\documentclass[10pt,journal,compsoc]{IEEEtran}

\usepackage{cite}

\usepackage{amsmath,amssymb,amsfonts}

\usepackage{graphicx}
\usepackage{textcomp}
\usepackage{xcolor}
\def\BibTeX{{\rm B\kern-.05em{\sc i\kern-.025em b}\kern-.08em
    T\kern-.1667em\lower.7ex\hbox{E}\kern-.125emX}}

\usepackage{balance}
\usepackage{amsthm}

\usepackage{booktabs} 

\usepackage{listings}
\usepackage{amsmath}
\usepackage{algorithm}
\usepackage{algpseudocode}

\usepackage{url}
\usepackage[lofdepth,lotdepth]{subfig}
\usepackage{multirow}
\usepackage{enumitem}

\newtheorem{challenge}{Challenge}
\newtheorem{theorem}{Theorem}

\newcommand{\eat}[1]{}
\newcommand{\mytilde}{\raise.17ex\hbox{$\scriptstyle\mathtt{\sim}$}}
\newlength{\oldtextfloatsep}

\newcommand\MYhyperrefoptions{bookmarks=true,bookmarksnumbered=true,
	pdfpagemode={UseOutlines},plainpages=false,pdfpagelabels=true,
	colorlinks=true,linkcolor={black},citecolor={black},urlcolor={black},
	pdftitle={Hihooi: A Database Replication Middleware for Scaling Transactional Databases Consistently},
	pdfsubject={},
	pdfauthor={Herodotos Herodotou},
	pdfkeywords={relationa databases, transaction processing, concurrency}}
\usepackage[\MYhyperrefoptions]{hyperref}

\begin{document}


\title{Hihooi: A Database Replication Middleware for Scaling Transactional Databases Consistently}

\author{Michael~A.~Georgiou, Aristodemos~Paphitis, Michael~Sirivianos, Herodotos~Herodotou
	\IEEEcompsocitemizethanks{
		\IEEEcompsocthanksitem Cyprus University of Technology, Limassol, Cyprus, 3036.\protect\\
		E-mails: mica.georgiou@edu.cut.ac.cy, am.paphitis@edu.cut.ac.cy, \protect\\
		michael.sirivianos@cut.ac.cy, herodotos.herodotou@cut.ac.cy}
}

\markboth{~~}%
{Georgiou \MakeLowercase{\textit{et al.}}: Hihooi: A Database Replication Middleware for Scaling Transactional Databases Consistently}

\IEEEtitleabstractindextext{%
	\begin{abstract}
		With the advent of the Internet and Internet-connected devices, modern business applications can experience rapid increases as well as 
variability in transactional workloads. Database replication has been employed to scale performance and improve availability 
of relational databases but past approaches have suffered from various issues including limited scalability, performance versus consistency 
tradeoffs, and requirements for database or application modifications. This paper presents Hihooi, a replication-based middleware system 
that is able to achieve workload scalability, strong consistency guarantees, and elasticity for existing transactional databases at a low cost.
A novel replication algorithm enables Hihooi to propagate database modifications asynchronously to all replicas at high speeds, while 
ensuring that all replicas are consistent. At the same time, a fine-grained routing algorithm is used to load balance incoming transactions to 
available replicas in a consistent way. Our thorough experimental evaluation with several well-established benchmarks shows how Hihooi 
is able to achieve almost linear workload scalability for transactional databases.

	\end{abstract}
	
}

\maketitle

\IEEEdisplaynontitleabstractindextext
\IEEEpeerreviewmaketitle

\ifCLASSOPTIONcompsoc
\IEEEraisesectionheading{\section{Introduction}\label{sec:intro}}
\else
\section{Introduction}
\label{sec:intro}
\fi

\IEEEPARstart{T}{he} recent explosion of data has led to the development of innovative systems for large-scale data processing \cite{parallelDBs-MR-FnTDB13}.
In the transactional systems arena, systems such as Google's Megastore \cite{megastore} and Spanner \cite{spanner-tocs13} were introduced for 
handling massive amounts of data, even across datacenter boundaries. However, the majority of transactional databases are smaller than 1TB in 
size \cite{hekaton-sigmod13}, indicating that excessive data scalability is a non-requirement for most small and medium enterprises (SMEs).
Rather, modern applications tend to experience both rapid growth and variability of users (and consequently application workload) due 
to the advent of the Internet and Internet-connected devices. Therefore, \textit{workload scalability}, i.e., the ability to handle 
increasing workload demands, as well as \textit{support for elasticity} to handle variations in those workloads, are critical 
for \textit{existing database instances}. 

\textit{NoSQL} technologies, such as MongoDB \cite{chodorow2013mongodb} and Cassandra \cite{cassandra-sigops10},
were explicitly designed to address scalability and elasticity requirements. By doing so, NoSQL systems sacrifice traditional consistency models
and the familiarity of SQL. Hence, they cannot replace existing transactional database systems.
More recently, a new class of systems has arisen, called \textit{NewSQL}, that offers similar scalable performance as NoSQL while still 
maintaining the ACID guarantees of a traditional database system \cite{nosql-newsql-survey}. NewSQL systems, however, are often highly 
optimized for a narrow set of use cases (e.g., MemSQL\cite{memsql} is tuned for clustered analytics) and require other compromises related to language support or transaction and workload handling capabilities (e.g., in VoltDB \cite{voltdb-engbul13}, the unit of transaction is a Java stored procedure). 
Finally, the evolution of cloud computing has led to several \textit{Database-as-a-Service} (\textit{DBaaS}) offerings (e.g., Amazon RDS, Azure SQL DB) that natively support 
scalability and elasticity in a pay-as-you-go model. 
Yet, SMEs are cautious in adopting them
due to the high costs associated with 
rewriting applications and retraining employees, as well as privacy and security concerns \cite{cloudcomp-survey-IS15}.

A typical approach to scaling an existing database system is to \textit{scale up}; i.e., to add more physical resources (e.g., memory, disks) 
to the server or upgrade to a higher-end server or a shared-disk database clustering solution (e.g., Oracle RAC \cite{oraclerac}).
Apart from being very expensive due to both hardware and software licensing costs, such solutions necessitate over-provisioning for peak 
and eventual volumes \cite{cecchet2008middleware}.
The alternative is to utilize a \textit{scale-out} approach, which can help reduce costs by hosting databases on multiple commodity hardware servers.
\textit{Data partitioning} (or \textit{sharding}) is one of the two main scale-out physical implementations, based on which the database data is partitioned 
and spread across all nodes \cite{scalable-sql-sigmodrec11}. While this approach does improve scalability (up to a point due to distributed 
transactions), it also requires expensive data migration and extensive manual physical design tuning for partitioning the data 
effectively \cite{calvin-sigmod12}.

The second scale-out approach is \textit{database replication}, which has been used for increasing performance and availability of databases 
under various requirements \cite{cecchet2008middleware}. This approach fully replicates data across all nodes and it comes in two forms, 
i.e., \textit{multi-master} and \textit{master-slave}. In the former, all replicas serve both read and write transactions but need 
explicit synchronization mechanisms in order to agree to a serializable execution order of transactions, so that each replica executes them 
in the same order \cite{c-jdbc, postgres-r, tashkent-2006}. Concurrent transactions might conflict, leading to aborts and thus limiting the 
system's scalability \cite{gray1996dangers}. 
In \textit{master-slave replication}, one primary copy handles all write operations while the other 
replicas process only read operations \cite{ganymed-middleware2004, middle-r-2005}. As long as the master node can handle the write workload, 
the system can scale linearly with the addition of more slave nodes \cite{cecchet2008middleware}. The biggest challenge here lies in the 
trade-off between performance and consistency of the overall system.

This paper presents \textit{Hihooi}, \textit{a replication-based master-slave middleware system that is able to achieve workload scalability, strong consistency, 
and elasticity 
for transactional databases}.
An existing database can readily become the master in a Hihooi deployment.
Replication is then used to increase the processing capacity of the system (which increases throughput), to spread the load across the nodes (which decreases latency), 
and to mask potential failures of individual nodes (which improves availability).
As a middleware system, Hihooi sits between the database engines and the client, offering a single database view and masking the complexity of the underlying replication.
Neither the database engines nor the clients need to be modified as long as the popular ODBC/JDBC drivers are used.
Load distribution, fault tolerance, and failure recovery
are all handled by Hihooi.

The novelty of Hihooi lies in its replication and transaction routing algorithms.
In particular, Hihooi replicates all write statements \textit{asynchronously} and applies them \textit{in parallel} at the replica nodes, while ensuring that all replicas remain consistent with the primary copy.
At the same time, a \textit{fine-grained transaction routing} algorithm ensures that all read transactions are load balanced to the replicas, while maintaining strong consistency semantics.
In particular, Hihooi supports \textit{global strong snapshot isolation}, explained and proved in Section \ref{sec:cc:levels}.
Finally, elasticity is achieved by supporting an easy and quick way to add and remove replicas from the cluster (partly due to the master-slave architecture).

Existing replication-based approaches fall short of achieving all of Hihooi's aforementioned desiderata.
Open-source solutions for replication are database-specific.
MySQL Cluster \cite{mysql-cluster} uses a synchronous replication mechanism which limits scalability.
Postgres-R \cite{postgres-r} integrates replica control into the kernel of PostgreSQL and utilizes special multicast primitives to propagate
low-level write operations to the replicas.
Middle-R \cite{middle-r-2005} allows all replicas to execute write transactions and uses a group communication system to determine a global commit order but requires database engine modifications for extracting and applying tuple-based modifications.
Finally, Ganymed \cite{ganymed-middleware2004} is a master-slave replication middleware similar to Hihooi but applies all changes serially at the replicas and offers only 
a coarse-grained load balancing of transactions.

In summary, the contributions of this paper are: 
\begin{enumerate}[leftmargin=*,itemsep=0pt]
 \item A new \textbf{database replication middleware architecture} for achieving workload scalability with strong consistency.
 \item A \textbf{statement replication algorithm} for applying writes in parallel while ensuring consistent replicas.
 \item Transaction- and statement-level \textbf{routing algorithms} for executing read transactions consistently and efficiently.
 \item An \textbf{extensive evaluation} showcasing the workload scalability that is attainable with Hihooi.
\end{enumerate}

\noindent
Section \ref{sec:overview} provides an overview of Hihooi that guides the rest of the paper until Section \ref{sec:evaluation}, which presents the experimental evaluation.
Section \ref{sec:related} presents related work and Section \ref{sec:conclusions} concludes the paper.

\section{Hihooi Overview}
\label{sec:overview}

Hihooi is a \textit{replication-based middleware} solution that aims at offering both workload scalability and strong consistency to enterprise databases.
As such, Hihooi employs \textit{master-slave replication}, a popular technique used to improve performance for transactional 
workloads \cite{cecchet2008middleware}. Transactions are categorized into \textit{write transactions} when at least one 
of the containing queries modifies the database (e.g., INSERT, UPDATE, DELETE SQL statements) and 
\textit{read transactions} otherwise. With master-slave replication, all write transactions are sent to the master node, 
denoted as {\it Primary DB}, while read transactions are directed to the slave nodes, denoted as {\it Extension DBs}.
As long as the Primary DB can handle all writes and the system propagates the writes to the Extension DBs efficiently, 
the system can scale linearly by adding more Extension DBs. However, achieving the dual goal of scalability 
and strong consistency introduces three core challenges addressed by the design choices of Hihooi.

\vspace{-2pt}
\begin{challenge}
\textbf{Replica Control:}
{How to efficiently and consistently propagate updates from Primary DB to Extension DBs.}
\end{challenge}
\vspace{-2pt}

\noindent
To ensure that the read transactions running at some Extension DB see a consistent view of the database, the 
replica must reflect a transaction-consistent snapshot of the data at the Primary DB; that is, the replica must reflect all 
data modifications of transactions (up to some transaction) executed at the Primary DB in the same order of execution.
To retain global system consistency, Hihooi captures the total order of transaction completions on the Primary DB and 
utilizes \textit{statement replication} (i.e., the write statements are replicated to the Extension DBs), while ensuring that 
each replica applies writes in the same order. The statement replication takes place \textit{asynchronously} in order to 
avoid delaying the write transactions executing at the Primary DB.

The conventional practice in database replication and hot standby deployments is to apply the writes serially at the slaves, 
even though the master processes them in parallel \cite{ganymed-middleware2004, wieck2005slonyi}.
In a heavily loaded production system, however, the lag between the master and a slave node can become significant \cite{cecchet2008middleware}.
Hihooi implements a novel algorithm for applying write transactions in parallel at the slaves, while maintaining strong consistency guarantees. 
The asynchronous propagation and the parallel execution of writes to the 
Extension DBs are described in Section \ref{sec:replication}.

\vspace{-2pt}
\begin{challenge}
\textbf{Concurrency Control:} 
How to efficiently and consistently route read transactions to  Extension DBs.
\end{challenge}
\vspace{-2pt}

\noindent
As a middleware layer between applications and database engines, Hihooi intercepts all incoming transactions and is tasked with
routing them either to the Primary DB or to one of the Extension DBs. Since all write transactions are always executed at the Primary DB, 
Hihooi can safely route there any read queries.
This tactic, however, goes against the primary goal of Hihooi to 
scale performance, which is maximized when the read queries are distributed to the available Extension DBs. The main issue here is that 
Extension DBs are not always up-to-date with the Primary DB due to the asynchronous propagation of write transactions to the Extension DBs. 
Hence, Hihooi needs an efficient approach for determining which Extension DBs are consistent with which incoming read queries.

The solution employed by Hihooi consists of three parts. 
First, for each incoming query $Q$, Hihooi utilizes a custom light-weight parser for extracting the tables, columns, or rows 
that are potentially modified by $Q$ (if $Q$ is a write query) or accessed by $Q$ (if $Q$ is a read query).
Second, Hihooi keeps track of the completed transactions that have been applied on each of the Extension DBs along with the transactions that 
are currently running on the Primary DB. Hence, Hihooi recognizes which tables, columns, or rows are up-to-date on each of the Extension DBs.
Finally, Hihooi employs a novel lightweight algorithm for checking which read queries are safe 
(from a consistency point of view) to execute on which Extension DBs. 
In the case where multiple Extension DBs can execute an incoming query, 
Hihooi will perform \textit{load balancing} and send the query to the least-loaded Extension DB. 
Hihooi is the first middleware system able to also do this for read queries that are part of multi-statement write transactions.
If no consistent Extension DB is found, then the 
system routes the request to the Primary DB, which is always consistent. 
The overall query interception and routing approach 
is detailed in Section \ref{sec:conc-control}.

\begin{figure}[t!]
	\centering
	\includegraphics[width=0.47\textwidth]{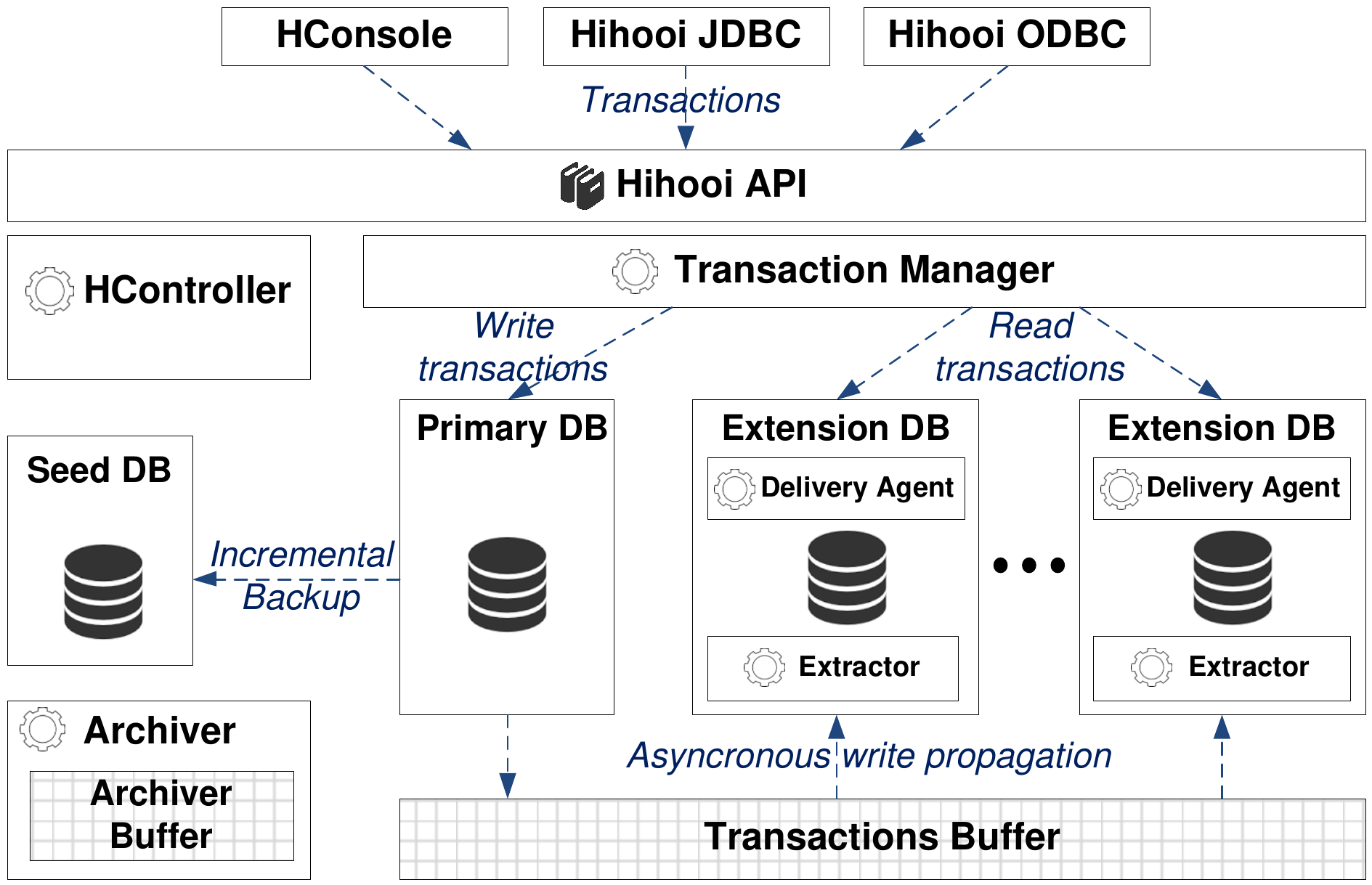}
	\caption{Hihooi Architecture}
	\label{fig:hihooi_architecture}
	\vspace{-3ex}
\end{figure}

\vspace{-2pt}
\begin{challenge}
\textbf{Backup and Scalability Management:} How to efficiently create a new Extension DB from a consistent backup and quickly bring it up-to-date.
\end{challenge}
\vspace{-2pt}

\noindent
Performing backups is a standard management operation for database systems in order to provide recovery from failures.
For a replicated database system, such as Hihooi, it is even more important because backups can be used to add new Extension DBs into the 
system \textit{without affecting the performance of the Primary DB or the existing Extension DBs}.
Hihooi periodically creates a backup, called \textit{Seed DB}, while being aware of exactly which transactions are contained in 
the backup and which ones must be re-executed to properly synchronize a new Extension DB.

A new Extension DB is created by cloning the Seed DB into a new node, followed by the execution of all write transactions 
missed since the creation of the backup. 
During the time needed to create and synchronize the new Extension DB, 
new write transactions may have been executed at the Primary DB.
Since Hihooi already allows for Extension DBs to fall behind and uses a smart query routing algorithm for executing queries correctly, 
it is not necessary to enact a global barrier to ensure consistency. Instead, as soon as the Extension DB is created and an initial set 
of write transactions has been executed, it can join the system and start executing read transactions, while concurrently 
applying the new write transactions. 
The backup and scalability management procedures are described in Section \ref{sec:scalability}.

\section{System Architecture}
\label{sec:architecture}

Figure \ref{fig:hihooi_architecture} depicts the Hihooi architecture, along with the core components and the flow of transactions through the system.
As a middleware system, Hihooi is positioned between the applications and the database engines.
The custom \textbf{Hihooi JDBC/ODBC Drivers} implement the \textbf{Hihooi API} and provide database-independent connectivity between the applications and Hihooi.
This approach requires the database driver to be replaced in the application but it does not require any other application code changes.
Internally, Hihooi uses JDBC drivers for interacting with the underlying database engines in order to execute the queries and to manage 
replication behind the scenes. Hence, Hihooi is not coupled to the database engines, thus supporting multiple vendors.
Currently, Hihooi supports multiple versions of the same engine running concurrently (which is important during database updates) 
and could support heterogeneous engines in the future. 
Finally, \textbf{HConsole} is an interactive console application that can be used 
for configuring and managing Hihooi, including adding and removing replicas, creating checkpoints, and executing queries. 

The \textbf{Transaction Manager} is responsible for intercepting all queries and sorting them into write and read transactions.
The write transactions are executed on the \textbf{Primary DB}, while the read transactions are load balanced to consistent \textbf{Extension DBs}.
Apart from managing the client sessions, the Transaction Manager oversees the available Extension DBs and keeps track of which write 
transactions they have applied.
Once a write transaction completes (either via \texttt{commit} or \texttt{rollback}), the transaction's statements are 
pushed into the \textbf{Transactions Buffer}, which is distributedly stored in memory using Memcached \cite{memcached-linuxj04}.
The Transactions Buffer acts as a highly available and fault tolerant propagation medium for all database modifications, which need to be applied asynchronously to the Extension DBs. 

\begin{table*}
	\centering
	\caption{Example write transactions on tables $\protect R(\underline{A1}, A2, A3, A4)$ and $\protect S(\underline{B1}, B2, B3, B4, B5)$ along with corresponding 
	write sets, read sets, affecting classes, and transaction state identifiers (TSIDs)}
	\label{tab:example-write-queries}
    \vspace{-1ex}
    \small
    {\setlength{\tabcolsep}{0.36em}
	\begin{tabular}{ |c|l|c|c|c|c|c|c|c|c| }
		\hline
		\multirow{2}{*}{TX} & \multirow{2}{*}{SQL Statement} & \multicolumn{3}{|c|}{Write Sets} & \multicolumn{3}{|c|}{Read Sets} & Affecting & \multirow{2}{*}{TSID} \\
        \cline{3-8}
		&  & Table & Column & Row & Table & Column & Row & Class  &  \\
		\hline
		$W_1$ & UPDATE R SET A2 = ?, A3 = ? WHERE A1 = 100 	& $R$ & $A2, A3$ & $A1=100$ & $R$ & $A1$ & $A1=100$ & RAS & 11 \\
		$W_2$ & UPDATE S SET B2 = ? WHERE B5 $>$ ? 			& $S$ & $B2$ & & $S$ & $B5$ &  & CAS & 12 \\
		$W_3$ & UPDATE R SET A3 = ?, A4 = ? WHERE A2 $<$ ? 	& $R$ & $A3, A4$ &  & $R$ & $A2$ & & CAS & 13 \\
		$W_4$ & DELETE FROM R WHERE A1 = 120 				& $R$ & $*$ & $A1=120$ & $R$ & $A1$ & $A1=120$ & RAS & 14 \\
		$W_5$ & UPDATE S SET B4 = ? WHERE B5 $<$ ? 			& $S$ & $B4$ &  & $S$ & $B5$ & & CAS & 15 \\
		\hline
	\end{tabular}
	}
    \vspace{-1ex}
\end{table*}

Each Extension DB node hosts two Hihooi services; the Extractor and the Delivery Agent.
The \textbf{Extractor} is responsible for fetching the new write transactions from the Transactions Buffer and applying them to the local database.
The Extractors implement a novel algorithm (discussed in Section \ref{sec:replication}) for 
executing the transactions in parallel, while respecting the order imposed by the transaction commit timestamps on the Primary DB. The \textbf{Delivery Agent} is 
responsible for executing the read-only queries routed to the local Extension DB and delivering the results set incrementally to the client when requested, to avoid creating an execution bottleneck at the Transaction Manager.

The \textbf{Archiver} has a dual role in Hihooi. First, it is responsible for initiating the incremental backups for creating the 
\textbf{Seed DB} based on the Primary DB, while keeping track with which transactions are included in the backup.
Hence, the Seed DB represents a consistent checkpoint of the Primary DB at some point in time.
Second, the Archiver periodically moves the write transactions that have been applied by all Extension DBs from the 
Transactions Buffer to its local and persistent \textbf{Archiver Buffer} in order to keep the memory usage of the Transactions Buffer bounded.
A new Extension DB is initialized using the Seed DB, followed by the application of the appropriate write transactions 
from the Archiver Buffer. Next, the Extension DB notifies the Transaction Manager that it can start serving read queries, 
while it starts applying the latest changes from the Transactions Buffer.
Finally, \textbf{HController} coordinates all system management operations, such as adding and removing Extension DBs.


\section{Database Replication}
\label{sec:replication}

Hihooi intercepts and redirects all incoming write transactions 
(i.e., transactions that modify the database) 
to the Primary DB for execution.
As soon as a transaction completes on the Primary DB, it must be propagated and executed on all Extension DBs, while preserving the completion order from the Primary DB. 
Before explaining our statement propagation and replication procedure in Section \ref{sec:repl:procedure}, we first introduce 
the notion of \textit{transaction read/write sets} in Section \ref{sec:repl:rwsets}. Finally, we discuss the benefits and practical considerations of our approach in 
Section \ref{sec:repl:practical}.

\subsection{Transaction Read/Write Sets}
\label{sec:repl:rwsets}

Transactions are naturally divided into \textit{single} and \textit{multi-statement}, depending on the number of SQL statements included in the transaction.
In most database engines, each SQL statement is considered to be a single transaction by default, and gets committed automatically when it completes execution.
Multi-statement transactions are either started with their first statement when automatic commit is disabled (and must be committed manually), or enclosed between specific 
keywords (e.g., \texttt{begin atomic ... end}). 
Hihooi follows the same conventions.
For ease of presentation, we discuss single transactions first, while multi-statement ones are presented if special handling is needed.

Each transaction $T$ will read and/or modify some tables in a database instance, defined as the \textbf{Table Read Set} and the \textbf{Table Write Set} of $T$, 
respectively. For example, transactions $W_1$, $W_2$, and $W_3$ shown in Table \ref{tab:example-write-queries} modify the respective tables $R$, $S$, and $R$; 
these tables form the corresponding table write sets.
Read/write sets allow us to reason about which transactions affect which tables. Thus, they allow us to effectively decide when to parallelize 
the execution of transactions on the Extension DBs (discussed in Section \ref{sec:repl:procedure}) and how to route read transactions efficiently 
(see Section \ref{sec:conc-control}). Suppose $W_1$--$W_3$ are executed on the Primary DB and committed in that order.
In general, we wish to execute $W_1$--$W_3$ on the Extension DBs in the same order to ensure the consistency of the replicas.
In this scenario, since $W_1$ and $W_2$ modify two different tables, we can execute them in parallel and let them commit in reverse order.
$W_3$, on the other hand, must execute after the completion of $W_1$ (as it modifies the same table $R$) in order to preserve consistency.

\begin{table}
	\centering
	\caption{Definitions of read/write sets for relational algebra operations. The read sets of write operations equal the corresponding read sets of 
		expression $E$}
	\label{tab:rwsets-rules}
	\vspace{-1ex}
	\small
	{\setlength{\tabcolsep}{3pt}
		\begin{tabular}{ |l|l|c|c|c| }
			\hline
			\multirow{2}{*}{Operation} & \multirow{2}{*}{Notation} & \multicolumn{3}{|c|}{Read Sets} \\
			\cline{3-5}
			&  & Table & Column & Row \\
			\hline
			Select & $\sigma_p{(R)}$ & $R$ & $A_i \in p$ & $(PK=?) \in p$ \\
			Project & $\Pi_{A_i}{(R)}$ & $R$ & $A_i$ & \\
			Union & $R \cup S$ & $R, S$ & $R.*, S.*$ & \\
			Set Difference & $R - S$ & $R, S$ & $R.*, S.*$ & \\
			Cartesian Pr. & $R \times S$ & $R, S$ & $R.*, S.*$ & \\
			Aggregation & $_{G_j}\mathcal{G}_{F_i(A_i)}{(R)}$ & $R$ & $G_j, A_i$ & \\
			\hline
			\multirow{2}{*}{Operation} & \multirow{2}{*}{Notation} & \multicolumn{3}{|c|}{Write Sets} \\
			\cline{3-5}
			&  & Table & Column & Row \\
			\hline
			Insert (tuple) & $R \leftarrow R \cup t$ & $R$ & $R.*$ & $(PK=?) \in t$ \\
			Insert & $R \leftarrow R \cup E$ & $R$ & $R.*$ & \\
			Delete & $R \leftarrow R - E$ & $R$ & $R.*$ & read set of $E$ \\
			Update & $R \leftarrow \Pi_{A'_i}{(E)}$ & $R$ & $A'_i$ if & read set of $E$ \\
			& & & $A'_i \neq R.A_i$ & \\
			\hline
			\multicolumn{5}{p{0.47\textwidth}}{{\footnotesize Symbols: $R$, $S$ = tables; $p$ = predicate; $A_i$ = attribute; $F_i$ = function; $G_j$ = group by attribute; $t$ = tuple; $E$ = relational algebra expression}} \\
		\end{tabular}
	}
	\vspace{-1ex}
\end{table}

Operating with table read/write sets constitutes a coarse-grained mechanism for reasoning about conflicting transactions.
Hence, we define two more levels of granularity for read/write sets.
The \textbf{Column Read/Write Sets} of a transaction $T$ denote the columns read/written by $T$.
Consider transaction $W_2$ from Table \ref{tab:example-write-queries}.
$W_2$ reads the column $S.B5$ (its column read set) and only updates $S.B2$ (its column write set).
Similarly, the column write set of $W_5$ is $\{S.B4\}$, which is disjoint from the column write set of $W_2$.
Hence, even though $W_2$ and $W_5$ modify the same table, they modify different columns and could be executed in parallel without affecting consistency.

\begin{table*}
	\centering
	\caption{Example read transactions on tables $\protect R(\underline{A1}, A2, A3, A4)$ and $\protect S(\underline{B1}, B2, B3, B4, B5)$ along with 
	the corresponding read sets, affecting classes, and consistent transaction state identifiers (TSIDs)}
	\label{tab:example-read-queries}
    \vspace{-1ex}
    \small
	\begin{tabular}{ |c|l|c|c|c|c|c| }
		\hline
		\multirow{2}{*}{TX} & \multirow{2}{*}{SQL Statement} & \multicolumn{3}{|c|}{Read Sets} & Affecting & Consistent \\
        \cline{3-5}
		&  & Table & Column & Row & Class & TSID \\
		\hline
		$R_1$ & SELECT * FROM R WHERE A2 $>$ ? 			& $R$ & $*$ & & TAS & 14 \\
		$R_2$ & SELECT A3, A4 FROM R WHERE A1 = 100		& $R$ & $A1, A3, A4$ & $A1=100$ & RAS & 13 \\
		$R_3$ & SELECT B2, B3 FROM S WHERE B5 $<$ ? 		& $S$ & $B2, B3, B5$ & & CAS & 12 \\
		$R_4$ & SELECT A1, B2, B3 FROM R JOIN S ON A1 = B2	& $R, S$ & $A1, B2, B3$ & & CAS & 14 \\
		\hline
	\end{tabular}
    \vspace{-0.5ex}
\end{table*}

Finally, the \textbf{Row Read/Write Sets} of a transaction $T$ denote the rows read/written by $T$ based on a primary key (PK) or a unique key.
For instance, transaction $W_1$ (see Table \ref{tab:example-write-queries}) updates the row in table $R$ for which $A_1=100$ ($A_1$ is the primary key 
of $R$), whereas $W_4$ deletes the row for which $A_1=120$.
Since $W_1$ and $W_4$ operate on different rows of the same table, they can also run concurrently without affecting consistency.
We restrict the row sets to include only primary or unique key equality predicates as those are simple to identify (using a basic query parser) and efficient to compare against each other.
The alternative would require reasoning with complex query-level semantics, whose overhead would potentially outweigh any of the performance benefits 
of concurrent execution.

Table \ref{tab:rwsets-rules} formalizes the creation of read/write sets based on fundamental relational algebra operations.
All read operations (i.e., select, project, union, set difference, Cartesian product, aggregation) result in table and column read sets that contain the accessed tables and columns, respectively. 
A select operation with a conjunctive equality predicate on the primary or unique key has a non-empty row read set.
Insert, delete, and update operations have both read and write sets.
The column write sets include all table columns for insert and delete operations, but only the modified columns for update operations.
The row write sets refer to the rows that are modified based on a primary or unique key (if applicable).
Finally, the read sets for write operations are based on the items accessed by their involved expressions.
Tables \ref{tab:example-write-queries} and \ref{tab:example-read-queries} list several write and read SQL statements along with their corresponding write and read sets.

Based on the scope by which an SQL statement affects a table $R$, we categorize it in one of three \textit{affecting classes}:
\begin{itemize}[leftmargin=*, itemsep=0pt]
\item \textbf{Row Affecting Statement (RAS)} when it modifies or accesses particular rows in $R$;
\item \textbf{Column Affecting Statement (CAS)} when it modifies or accesses some columns of $R$;
\item \textbf{Table Affecting Statement (TAS)} when it modifies or accesses all columns of $R$.
\end{itemize}
\noindent These class definitions will be utilized by our algorithms presented later in Sections \ref{sec:repl:procedure} and \ref{sec:conc-control}.
Tables \ref{tab:example-write-queries} and \ref{tab:example-read-queries} also include the affecting classes for each example transaction.

All aforementioned definitions are easily extended to a multi-statement transaction $T_m$.
The table read set of $T_m$ is simply the union of all table read sets of the individual SQL statements in $T_m$.
The same applies for the table/column/row read/write sets of $T_m$.
As for the affecting class of $T_m$, the following rules apply:
(i) if all statements in $T_m$ are ``RAS'' for table $R$ then $T_m$ is a ``RAS'' for $R$;
(i) if all statements in $T_m$ are ``CAS'' for $R$ then $T_m$ is a ``CAS'' for $R$;
(iii) otherwise, $T_m$ is a ``TAS'' for $R$.
Finally, DDL statements are handled as ``TAS'' write statements.

\subsection{Statement Replication Procedure}
\label{sec:repl:procedure}

For each transaction $T$ intercepted by Hihooi, a \textit{Transaction State} (or \textit{TState}) is built and maintained at the Transaction Manager. A TState contains the following:
\begin{enumerate}[leftmargin=*, itemsep=0pt]
  \item a TState identifier (\textit{TSID}) that uniquely identifies $T$; 
  \item the \textit{SQL write statements} of $T$ in the order of execution;
  \item the read/write sets of each statement and the overall $T$;
  \item the total execution time of $T$ on the Primary DB; and
  \item the completion operation: \texttt{commit} or \texttt{rollback} (the replication of failed transactions is explained in Table \ref{tab:practical-issues}).
\end{enumerate}

\noindent
\textbf{TSID Generation:}
TSID is a unique sequential number given to the TState of $T$ once $T$ commits or rollbacks on the Primary DB.
The purpose of the TSIDs is to capture the order of transaction completions on the Primary DB, which is determined by the \textit{transaction commit timestamps} as recorded by the underlying database system.
In order to ensure the correct ordering, the Transaction Manager (TM) performs the following steps after $T$ completes:
(1) the TM obtains the commit timestamp of $T$ and the commit timestamps of any other concurrent transactions that have issued a commit request but have not received a response yet;
(2) the TM issues TSIDs for these transactions in the same order as the commit timestamps.
Typically, $T$'s timestamp is the lowest and it simply receives the next TSID number.
Occasionally though, commit responses are received out of order (up to 2\% of the times in our most write-intensive experiments) due to multi-thread scheduling in the TM or network delays.
Hence, the procedure above is necessary to ensure that TSIDs are given in the same order induced by the commit timestamps.
The TState must be given a TSID 
before it can be fetched and replayed by the Extension DBs.

Each Extension DB node hosts an Extractor service, which is responsible for receiving the completed TStates from the Transactions Buffer and executing them on the 
local Extension DB. The goal of the Extractor is to ensure that the local database replica is consistent with the Primary DB.
Executing the transactions from the TStates in the serial order imposed by the TSIDs is a sufficient condition to achieve consistency.
However, it is very inefficient and can cause the Extension DBs to fall significantly behind the Primary DB, especially in times of heavy write 
load (given the parallel execution at the Primary DB). Hence, it is crucial for the Extractors to execute in parallel as many transactions as 
possible while maintaining correct consistency semantics.

As it was alluded in Section \ref{sec:repl:rwsets}, the read/write (R/W) sets of the transactions are the backbone for our parallel execution algorithm. 
Specifically, the R/W sets of two write transactions can be used to determine whether the transactions affect the same data items in the database, 
as shown in Algorithm \ref{algo:areindependent}.
If they don't affect the same items, we say they are \textbf{independent}. 
When the table R/W sets of two write transactions are disjoint, 
they are independent as they modify different tables (lines\#2-4). 
Otherwise, we need to check which columns and/or rows are modified by the two transactions, but only for 
the commonly modified tables (line\#6). If the two transactions both belong to class ``CAS'' for a table $t$ (i.e., they affect some columns of $t$) but do not 
modify the same columns, then they are independent for $t$ (lines\#8-11).
Similarly, if the two transactions both belong to class ``RAS'' for a table $t$ (i.e., they affect some rows of $t$) but don't modify the same rows, then they are 
independent for $t$ (lines\#12-15). 
If either of the above two conditions holds for all common tables, then the transactions are independent and it is safe to execute them concurrently.

One important property of our R/W set definitions is their \textit{cumulative} nature.
That is, if we take the union of the R/W sets of multiple statements, we get R/W sets of a multi-statement transaction with the same correct semantics, 
as explained in Section \ref{sec:repl:rwsets}.
With the same reasoning, we can combine the read/write sets of two or more multi-statement transactions that are running in parallel to build a transaction state 
that represents all running statements as if they were one bigger multi-statement transaction.
For example, suppose transactions $W_1$ and $W_2$ from Table \ref{tab:example-write-queries} are executed together by an Extractor.
Then, we can define a \textit{running transaction state} that includes the merged states of $W_1$ and $W_2$. 
The table write set of this new state would include tables $R$ and $S$, meaning both tables are currently being modified.
This combined state allows us to avoid checking whether a new transaction is independent with each currently running transaction.
Instead, we only need to check whether the new transaction is independent with the combined running transaction state.

\setlength{\textfloatsep}{3ex}

\begin{algorithm}[t!]
	\small
	\caption{\small Check independence between 2 write transactions}
	\label{algo:areindependent}
	\begin{algorithmic}[1]
		\Function{areIndependent}{$ts_1$, $ts_2$}
		\If {$ts_1.table\_w\_set \cap ts_2.table\_rw\_set = \varnothing \And$ \\
			\hskip2.5em $ts_1.table\_rw\_set \cap ts_2.table\_w\_set = \varnothing$}
		\State \Return true
		\EndIf
		\For{\textbf{each} t \textbf{in} 
			$ts_1.table\_rw\_set \cap ts_2.table\_rw\_set$}
		\State \textbf{bool} independent $\gets$ false
		\If {$ts_1.class[t]=CAS \And ts_2.class[t]=CAS \And$ \\ 
			\hskip4em $ts_1.col\_w\_set[t] \cap ts_2.col\_rw\_set[t] = \varnothing \And$ \\ 
			\hskip4em $ts_1.col\_rw\_set[t] \cap ts_2.col\_w\_set[t] = \varnothing$}
		\State independent $\gets$ true
		\ElsIf {$ts_1.class[t]=RAS \And ts_2.class[t]=RAS \And$ \\ 
			\hskip4em $ts_1.row\_w\_set[t] \cap ts_2.row\_rw\_set[t] = \varnothing \And$ \\ 
			\hskip4em $ts_1.row\_rw\_set[t] \cap ts_2.row\_w\_set[t] = \varnothing$}
		\State independent $\gets$ true
		\EndIf
		\State \textbf{if} independent == false \textbf{then return} false \textbf{end if}
		\EndFor
		\State \Return true
		\EndFunction
	\end{algorithmic}
	{\footnotesize
		Notation: $*\_w\_set$ = (table $|$ column $|$ row) write set;
		$*\_rw\_set$ = union of (table $|$ column $|$ row) read and write sets}
\end{algorithm}

Algorithm \ref{algo:parallelexecution} shows the two functions that constitute the parallel execution algorithm employed by the Extractor.
The new transaction states are received by the Extractor in the order imposed by the Primary DB execution.
For each new transaction state $tsNew$, the Extractor checks if it is independent from both the running and waiting states, i.e., the combined states of 
the already running and waiting transactions, respectively (lines\#5-6).
If it is, $tsNew$ can be executed in parallel with the already running transactions, after its state is merged with the state of the running transactions (lines\#7-8).
Otherwise, $tsNew$ must wait in the queue and its state must update the waiting state (lines\#10-11).
It is important to check $tsNew$ against the waiting transactions because a conflict indicates that an already waiting transaction will 
modify a data item that $tsNew$ will affect, and these changes must occur in order.

\begin{algorithm}[t!]
  \small
  \caption{\small Parallel transaction execution at Extension DBs}
  \label{algo:parallelexecution}
  \begin{algorithmic}[1]
    \State $runningState$ \Comment{combined state of running transactions}
    \State $waitingState$ \Comment{combined state of waiting transactions}
    \State $waitQueue$ \Comment{FIFO queue with waiting transaction states}
    \Function{onNewTransaction}{$tsNew$}
    \If {areIndependent($runningState$, $tsNew$) $\And$ \\ 
         \hskip2.4em areIndependent($waitingState$, $tsNew$)}
        \State $runningState$.merge($tsNew$)
        \State execute($tsNew$)
    \Else
        \State $waitingState$.merge($tsNew$)
        \State $waitQueue$.enqueue($tsNew$)
    \EndIf
    \EndFunction
    \Function{onTransactionComplete}{$tsOld$}
    \State $runningState$.remove($tsOld$)
    \While {$waitQueue$.isNotEmpty() $\And$ \\ 
         \hskip2.4em areIndependent($runningState$, $waitQueue$.peek())}
        \State $tsRun \leftarrow waitQueue$.dequeue()
        \State $waitingState$.remove($tsRun$)
        \State $runningState$.merge($tsRun$)
        \State execute($tsRun$)
    \EndWhile
    \EndFunction
  \end{algorithmic}
\end{algorithm}

Suppose the five write transaction from Table \ref{tab:example-write-queries} must be executed at an Extension DB in that order.
Transactions $W_1$ and $W_2$ will execute in parallel as they modify different tables, while $W_3$ is placed in the wait queue since it conflicts with $W_1$.
Even though $W_4$ is independent from the two running transactions ($W_1$ and $W_2$), it is not independent from the waiting $W_3$ and, hence, will also be 
placed in the wait queue. $W_5$ can also run in parallel as it modifies a different table than $W_1$ and a different column than $W_2$.

\begin{table*}
	\centering
	\caption{Practical issues and resolutions for statement replication}
	\label{tab:practical-issues}
    \vspace{-1ex}
    \small
	\begin{tabular}{ |p{4.8cm}| p{12.5cm}| }
	\hline
	   \textbf{Practical Issue} & \textbf{Resolution}  \\
	\hline
		Database sequences, used to generate unique or auto-incremented keys, are non-transactional objects
		& 
        Failed transactions are also executed on the Extension DBs to increment any sequences consistently.
        Read transactions that perform sequence operations are treated as write transactions (i.e., executed on the Primary DB and replicated to the Extension DBs).
    \\
    \hline
        Time-related macros such as \texttt{now} or \texttt{current\_timestamp}
		& 
		Query rewriting techniques are used to replace a macro with an actual value that will be common to all replicas.
	\\
	\hline
		User-defined functions in write transactions
		& 
        Non-deterministic functions are executed once and their return values are used in the calling statements.
        Deterministic functions are left as is.
	\\
	\hline
		Stored procedures and database triggers
		& 
		The R/W sets are extracted from deterministic procedures and triggers upon their definition. For deterministic procedures with non-SQL 
		definitions, the DB admin must provide their table R/W sets during system configuration. Non-deterministic procedures/triggers are currently 
		not supported.
	\\
	\hline
	\end{tabular}
    \vspace{-1ex}
\end{table*}

When a running transaction completes execution, its state is removed from the running state (line\#15).
Next, the wait queue is iterated, checking if the next waiting transaction is independent from the running transactions (lines\#16-17).
If it is, its state is moved from the waiting to the running state and then submitted for execution (lines\#18-21).
In our running example, when $W_1$ completes execution, $W_3$ can then execute, followed by $W_4$.

Even though we refer to the read/write sets as ``sets'', they are internally implemented using \textit{hash tables}, where the key is the data item (i.e., table, column, or row) and the value is a counter to track how many query statements access that data item.
In addition, the column and row read/write sets are maintained separately per database table (also using hash tables), to facilitate their direct use in Algorithms \ref{algo:areindependent} and \ref{algo:addToIndexes} (presented in Section \ref{sec:cc:transaction}).
The merging of two transaction states 
is a straightforward process.
For each corresponding read/write set, the underlying hash tables are merged as follows: if the two hash tables contain the same key, the associated counters are added together; otherwise the entries are simply put in the resulting table.
The deletion of a transaction state from a previously merged state involves decreasing the counters kept for each read/write set.
If a counter reaches zero, then the corresponding entry is deleted from the hash table.
Hence, all operations on read/write sets are very efficient to implement in practice.

Finally, the Extractor is responsible for notifying the Transaction Manager with the latest applied TSID in sequential order without gaps.
Suppose the transactions complete in the order $W_1$, $W_3$, $W_2$.
When $W_1$ completes, its TSID is reported but when $W_3$ completes nothing is reported. 
Once $W_2$ completes, the TSIDs of $W_2$ and $W_3$ are reported. 
Hence, the Transaction Manager is aware of up to which transaction has been replayed on each Extension DB in sequential order.
This information is stored in a simple hash table that maps the Extension DBs to their latest applied TSID.

\subsection{Benefits and Practical Considerations of Statement Replication}
\label{sec:repl:practical}

Hihooi employs \textit{statement replication} to ensure consistency in the replicas; i.e., it replicates and executes all write statements on the Extension DBs.
The alternative approach would be to use \textit{row-based replication}, which entails capturing the modified table rows on the Primary DB and replicating them to 
the Extension DBs \cite{postgres-r, middle-r-2005, tashkent-2006}.
One method for achieving row-based replication is to integrate the middleware with the underlying database engine for extracting and adding table rows.
This limits the ability of the middleware to use different database engines (or even different versions of the same engine) \cite{qin2017}.
Another method is to (i) declare triggers on every table for extracting the modifications and (ii) use a complex mechanism for applying the changes to the 
replicas. A serious drawback here is the performance overhead introduced on the primary database from the multiple trigger executions.

Hihooi avoids the aforementioned limitations of row-based replication by using statement replication.
Hence, it is capable of supporting multiple, unmodified database engines as well as preventing any unnecessary overheads at the Primary DB.
In addition, update and delete statements that affect multiple rows are very efficient to replicate as only the SQL statements are propagated to the replicas \cite{qin2017}.
Finally, capturing statements is the basis for extracting the R/W sets, which are used to improve the performance of both 
the replication and the query routing procedures.

The main practical issues concerning statement replication arise due to \textit{non-deterministic queries}, i.e., queries that may not produce the same result even when executed on the same database state.
Statement replication requires that the execution of a write statement has the same effect on the Primary DB as on the Extension DBs.
However, an SQL statement could legitimately produce different results on different replicas if, for example, it referenced sequences, or used the current timestamp, 
or invoked a non-deterministic function (e.g., RAND). 
Hihooi resolves such issues by (i) performing on-the-fly query rewriting before submitting the queries for execution; and 
(ii) replicating all, even failed, transactions. Table \ref{tab:practical-issues} lists the main practical issues along with Hihooi's resolutions in the present  implementation.
Currently, Hihooi does not support the small set of non-deterministic  procedures and triggers.

\eat{ 
In terms of replication scope, Hihooi supports \textit{Schema-} and \textit{Table-based} replication.
In the former, Hihooi replicates the entire database schema (including tables, indexes, etc.) and all changes resulted from the execution of 
write transactions on the Primary DB. 
In table-based replication, the system replicates only a specified set of tables along with their indexes and other supporting 
data structures. Similarly, only changes that affect those tables are replicated to the Extension DBs.
Therefore, only read transactions that exclusively access those tables are distributed to the Extension DBs; all others execute on the Primary DB.
}

\setlength{\textfloatsep}{\oldtextfloatsep}

\section{Concurrency Control} 
\label{sec:conc-control}

As explained in Section \ref{sec:replication}, all write transactions are executed on the Primary DB, are given a sequential TSID upon completion, 
and are replicated to the Extension DBs. An Extension DB is considered \textit{consistent} if it has replicated all transactions up to the latest transaction 
(which has the largest TSID) that was executed on the Primary DB.
Read transactions can safely be routed either to the Primary DB or to any consistent Extension DB for execution.
However, the asynchronous replication of write transactions to the Extension DBs can result in a lag between the Primary DB and the Extension DBs.
In such a scenario, read transactions must either wait for at least one Extension DB to become consistent or be redirected to 
the Primary DB. The first option introduces latency delays for the read transactions, while the second further increases the load on the Primary DB.
In either case, performance and scalability can suffer.

Hihooi implements a novel \textit{transaction-level} routing and load balancing algorithm that utilizes read/write sets for directing transactions to Extension DBs, 
even if they are not consistent with the Primary DB.
The key idea is that it is safe to route a read transaction $T$ to an Extension DB if the tables (or columns/rows) accessed by $T$ will not be modified by the write 
transactions that have yet to execute on the Extension DB (see Section \ref{sec:cc:transaction}).
Further, Hihooi can perform an even finer-grained load balancing by directing individual read statements from within multi-statement write transactions to Extension DBs.
To the best of our knowledge, \textit{Hihooi is the first replication-based middleware system to offer statement-based routing}, while respecting transaction 
boundaries and maintaining consistency (see Section \ref{sec:cc:statement}).

\subsection{Transaction-level Load Balancing}
\label{sec:cc:transaction}

The goal of transaction-level load balancing is to direct read transactions to Extension DBs that are consistent with the Primary DB but only with regards to the data 
each read transaction will access.
In order to achieve this efficiently, Hihooi needs quick access to the tables, columns, or rows accessed by the read transactions as well as to which tables, 
columns, or rows are up-to-date on each Extension DB.
The former is achieved using the transaction read sets, while the latter using the TSIDs of the completed write transactions and a set of hash indexes 
maintained by the Transaction Manager.
In particular, three hash indexes are used for separately mapping tables, columns, and rows to the latest write transaction that modified them.
Hence, the indexes can be used to find the transaction after which the replica is consistent with regards to specific tables, columns, or rows.

Once a write transaction $T_w$ completes, its write sets are used to update the three indexes, as shown in Algorithm \ref{algo:addToIndexes}.
All tables referenced in the table write set of $T_w$ are added into the \textit{Tables Hash Index} ($TIndex$) and mapped to the transaction state identifier 
(TSID) of $T_w$ (line\#2). This action indicates that the latest transaction to update those tables is $T_w$.
Next, the modification of the other two indexes depends on the affecting class of $T_w$ for each table.
In particular, if $T_w$'s class is ``TAS'' or ``CAS'', then all columns in $T_w$'s column write set are added into the \textit{Columns Hash Index} ($CIndex$) and mapped 
to $T_w$'s TSID (lines\#4-5). Otherwise, all rows in $T_w$'s row write set are added into the \textit{Rows Hash Index} ($RIndex$) and mapped to $T_w$'s TSID (lines\#6-7).
Row entries in RIndex that have been applied to all replicas are periodically pruned to keep the index size bounded.

Consider the five write transactions of our running example shown in Table \ref{tab:example-write-queries}.
After their execution on the Primary DB, the content of the three hash indexes is shown in Table \ref{tab:content-indexes}.
Each entry (of any index) shows the last TSID that modified that particular item.
For example, table $S$ was last modified by transaction $W_5$ with TSID=15, while column $S.B2$ was last modified by $W_2$ with TSID=12.

\setlength{\textfloatsep}{3ex}

\begin{algorithm}[t!]
  \small
  \caption{\small Update the Transaction Manager hash indexes after executing a write transaction}
  \label{algo:addToIndexes}
  \begin{algorithmic}[1]
    \Function{updateIndexes}{$ts$}
    \State $TIndex$.multiPut($ts.table\_write\_set$, $ts.TSID$)
    \For{\textbf{each} t \textbf{in} $ts.table\_write\_set$}
        \If {$ts.class[t]=TAS$ $||$ $ts.class[t]=CAS$}
            \State $CIndex$.multiPut($ts.col\_write\_set[t]$, $ts.TSID$)
        \ElsIf {$ts.class[t]=RAS$}
            \State $RIndex$.multiPut($ts.row\_write\_set[t]$, $ts.TSID$)
        \EndIf
    \EndFor
    \EndFunction
  \end{algorithmic}
\end{algorithm}

The last step in the transaction-level load balancing is to determine which Extension DBs are consistent for running an incoming read transaction $T_r$.
Algorithm \ref{algo:findConsistentTSID} shows how the indexes can be used for finding the TSID of the last transaction that modified any of the data items 
accessed by $T_r$. 
For each table $t$ to be accessed by $T_r$, we utilize the affecting classes and read sets for 
guiding our algorithm, assuming $t$ has been modified before (lines\#3-4).
If $T_r$ is a ``$TAS$'' for table $t$ (i.e., it will access all columns of $t$), then a single lookup on the $TIndex$ is enough to find the latest TSID (lines\#5-6).
If $T_r$ is a ``$CAS$'' for $t$ (i.e., it will access some specific columns of $t$), then we need to (i) lookup the $CIndex$ to find the latest TSID among all 
columns in $T_r$'s column read set and (ii) search the $RIndex$ to find the largest TSID from all rows affecting $t$ (lines\#7-9). 
We search for the rows as well since any row modification can potentially modify any column.
Finally, if $T_r$ is a ``$RAS$'' for $t$ (i.e., it will access some specific rows of $t$), then we also lookup both $RIndex$ and $CIndex$ (lines\#10-12).
Overall, we return the largest TSID found from all lookups across all tables to ensure consistency.
Any Extension DB that has replicated at least that TSID can be used for executing $T_r$.

\begin{algorithm}[t]
  \small
  \renewcommand\algorithmicindent{0.9em}
  \caption{\small Find the latest consistent TSID for a read transaction on the Transaction Manager}
  \label{algo:findConsistentTSID}
  \begin{algorithmic}[1]
    \Function{findLatestConsistentTSID}{$ts$}
    \State $tsid = 0$
    \For{\textbf{each} t \textbf{in} $ts.table\_read\_set$}
        \State \algorithmicif\ \textbf{not} $TIndex$.contains($t$)\ \algorithmicthen\ \textbf{skip iteration}
        \If {$ts.class[t]=TAS$}
            \State $tsid=$ max\{$tsid$,  $TIndex$.lookup($t$)\}
        \ElsIf {$ts.class[t]=CAS$}
            \State $tsid=$ max\{$tsid$,  $CIndex$.lookup($ts.col\_read\_set[t]$)\}
            \State $tsid=$ max\{$tsid$,  $RIndex$.maxValue($t$)\}
        \ElsIf {$ts.class[t]=RAS$}
            \State $tsid=$ max\{$tsid$,  $RIndex$.lookup($ts.row\_read\_set[t]$)\}
            \State $tsid=$ max\{$tsid$,  $CIndex$.lookup($ts.col\_read\_set[t]$)\}
        \EndIf
    \EndFor
    \State \Return $tsid$
    \EndFunction
  \end{algorithmic}
\end{algorithm}

\begin{table}[t]
	\centering
	\caption{The content of the Tables, Columns, and Rows Hash Indexes after executing the Table \ref{tab:example-write-queries} transactions}
	\label{tab:content-indexes}
    \vspace{-0.5ex}
    \small
	\begin{tabular}{ |c||c||c| }
		\hline
		$TIndex$ & $CIndex$ & $RIndex$ \\
		\hline
		$R \rightarrow 14$ & $A3 \rightarrow 13 $ & $(A1=100) \rightarrow 11$ \\
		                   & $A4 \rightarrow 13 $ & $(A1=120) \rightarrow 14$ \\
		$S \rightarrow 15$ & $B2 \rightarrow 12 $ &  \\
		                   & $B4 \rightarrow 15 $ &  \\
		\hline
	\end{tabular}
    \vspace{-1ex}
\end{table}

Table \ref{tab:example-read-queries} lists some example read transactions along with their consistent TSID based on the indexes' content in Table \ref{tab:content-indexes}.
Consider transaction $R_3$ that accesses columns $B2$, $B3$, and $B5$ of table $S$.
Based on Table \ref{tab:content-indexes}, only the relevant column $S.B2$ has been modified by transaction with TSID=12.
Hence, $R_3$ can execute on any Extension DB that has applied transactions with TSID=12 or higher.

\subsection{Statement-level Load Balancing}
\label{sec:cc:statement}

Master-slave replication dictates that all transactions that modify the database, including multi-statement ones, must be executed on the master first.
However, multi-statement write transactions may contain several read SQL statements, all of which are now executed on the Primary DB.
Some of these reads could potentially be executed on Extension DBs without violating atomicity or consistency constraints and, hence, increase the scalability of the entire system.

The premise is that a read statement within a multi-statement write transaction $T_m$ that is independent of its preceding write statements in $T_m$ can be 
safely executed on a consistent Extension DB.
This premise does not hold for serializable execution, but does hold for Snapshot Isolation, which is the default consistency level of Hihooi (see Section \ref{sec:cc:levels}), because the read still sees a consistent snapshot of the database.
Algorithms \ref{algo:areindependent} and \ref{algo:findConsistentTSID} can be used to efficiently check independence and find a consistent Extension DB, respectively.
In particular, when the write statements of $T_m$ are executed on the Primary DB, a running state is kept by the Transaction Manager (similar to the running state 
kept by the Extractors described in Section \ref{sec:repl:procedure}).
When a read statement arrives in $T_m$, Hihooi checks if it is independent from the running state (recall Algorithm \ref{algo:areindependent}).
If so, Algorithm \ref{algo:findConsistentTSID} is used as in Section \ref{sec:cc:transaction} to find the latest consistent TSID, and thus, the available Extension DBs for execution.
When the read is not independent from the previous writes, or no consistent replica is found, it 
is executed on the Primary DB.

\eat{
In the presence of multiple consistent Extension DBs, a pluggable load balancing policy is used to select one of them for executing the read statement.
The current policy simply selects the Extension DB with the lowest number of active connections but any load balancing technique could be used.
}

\subsection{Consistency Levels}
\label{sec:cc:levels}

Most database engines (e.g., PostgreSQL, Oracle, DB2) use \textit{snapshot isolation} (\textit{SI}) for enforcing consistency 
\cite{cecchet2008middleware}.
With SI, each transaction operates on its own copy of data (a snapshot), allowing read transactions to complete without blocking.
Similarly, database replication research has been focusing on SI and its variants, such as Generalized SI, Strong SI, and Weak SI \cite{middleware-si-sigmod05}.
Hihooi works over a set of SI-based database replicas and offers the illusion of a single SI database to the client.
Hence, it provides a form of \textbf{Global Strong Snapshot Isolation (GSSI)} \cite{middleware-si-sigmod05}.


We follow concepts introduced in \cite{Berenson_1995, middleware-si-sigmod05, lazy-repl-SI-vldb06, isolation-levels-icde00} in order to formalize the notion of GSSI 
in replicated systems and develop a direct proof of its support by Hihooi.
According to SI, as introduced in \cite{Berenson_1995}, the system assigns a transaction $T$ a start timestamp $s(T)$ at the beginning of its execution, before performing any read or write operations.
$T$ will always read data from a snapshot of the (committed) data as of $s(T)$.
In particular, writes performed by any transaction $T'$ that commits
before $s(T)$ will be visible to $T$. 
On the other hand, writes performed by any transaction $T'$ that commits after  $s(T)$ will not be visible to $T$. 
SI also requires that each transaction $T$ be able to see its own writes, even though the writes occurred after $s(T)$.
After finishing its operations, $T$ is assigned a commit timestamp, $c(T)$, such that $c(T)$ is more recent than any start or commit timestamp assigned to any transaction.
$T$ commits only if all other transactions $T'$ that committed during the lifespan of $T$ (i.e., $s(T) < c(T') < c(T)$) did not modify any data that $T$ has also written.
Otherwise, $T$ is aborted so as to prevent lost updates.
Note that two transactions $T_1$ and $T_2$ are called \textit{concurrent} if their lifespan intervals $[s(T_1), c(T_1)]$ and $[s(T_2), c(T_2)]$ overlap.

According to the original definition of SI, the system can choose $s(T)$ to
be any time less than or equal to the actual start time of $T$.
Hence, $T$ can see any snapshot earlier than its start timestamp and not necessarily the latest one.
This relaxed version of SI is called \textit{Weak SI} in \cite{lazy-repl-SI-vldb06}.
With \textit{Strong SI}, a transaction $T_2$ that starts after a committed
transaction $T_1$ is guaranteed to see a committed database state
that includes the effects of $T_1$.
In other words, $T_2$ will see the latest snapshot of the database state.
Most current database systems (including PostgreSQL) and research prototypes \cite{postgres-r-si-icde05, middle-r-2005, ganymed-middleware2004} offer Strong SI.
Finally, the qualifier \textit{`global'} indicates that the definition of Strong SI applies to the distributed system as a whole and not to the individual database replicas.

Summarizing, a transaction history in a replicated database system satisfies Global Strong SI if
its committed transactions satisfy the following two conditions: 
\begin{enumerate}
	\item Read operations in any transaction $T$ see the database in the state after the last commit before $s(T)$. Read operations in $T$ also see the data values that were last written by $T$ itself;
	\item Concurrent transactions do not modify the same data objects in the database.
\end{enumerate}

\begin{theorem}
	If each underlying database system in the replicas guarantees Strong SI, the Hihooi guarantees Global Strong SI.
\end{theorem}

\begin{proof}
Given a set of transactions to be executed with Hihooi, we need to show that their transaction history will satisfy the two conditions of Global Strong SI noted above.

\noindent
\textbf{Condition 1:}
Suppose $T$ is a read transaction arriving for execution in Hihooi.
Algorithm \ref{algo:findConsistentTSID} will find the TSID of the last write transaction that modified any of the data to be accessed by $T$ (recall Section \ref{sec:cc:transaction}).
Then, $T$ will be routed to any Extension DB that replicated at least that TSID, guaranteeing that $T$ will see the latest relevant state.
When no such Extension DB is found, $T$ is routed to the Primary DB, which is always up to date and offers Strong SI by itself.
If $T$ is a write transaction, then it will be executed on the Primary DB. Since all write transactions always execute on the Primary DB and the Primary DB guarantees Strong SI locally, any read operations in $T$ will see the latest database state. 
With statement-level load balancing (recall Section \ref{sec:cc:statement}), read statements in $T$ that are independent of their preceding write statements in $T$, can be routed to Extension DBs.
Since the routing algorithm is the same as the one used for read-only transactions, these read statements will see the latest relevant state as explained above.
Read statements in $T$ that access data written by previous write statements in $T$ are sent to the Primary DB and, hence, will see the data values that were last written by $T$.

\noindent
\textbf{Condition 2:}
Since all write transactions are always executed on the Primary DB, which offers Strong SI, no concurrent transactions can modify the same data and commit successfully on the Primary DB.
On the Extension DBs, transactions that modify the same data are never run concurrently per Algorithm \ref{algo:parallelexecution} (recall Section \ref{sec:repl:procedure}).
Only independent write transactions are ever executed in parallel on the Extension DBs, guaranteeing that concurrent transactions do not modify the same data in the database.
\end{proof}

By controlling the replication and routing mechanisms, Hihooi can 
offer three additional consistency levels
at the granularity of a database session:
\begin{enumerate}[leftmargin=*, itemsep=0pt]
\item \textbf{Weak SI}: Write transactions are asynchronously executed on the Extension DBs and read transactions are sent to any Extension DB regardless of their consistency.

\item \textbf{Replicated SI with Primary Copy (RSI-PC)}: Write transactions are asynchronously executed on the Extension DBs and read transactions are sent to 
any Extension DB that is fully consistent with the Primary DB (but waits if none is available). RSI-PC, another form of GSSI, is implemented by the middleware 
Ganymed \cite{ganymed-middleware2004}.

\item \textbf{One-copy Serializability (1SR)}: Write transactions are synchronously executed on all Extension DBs and read transactions are sent to any Extension DB. 
1SR is the default consistency level of middleware C-JDBC \cite{c-jdbc}.
\end{enumerate}

\setlength{\textfloatsep}{\oldtextfloatsep}

\section{Scalability Management}
\label{sec:scalability}

Performing backups and adding/removing Extension DBs are important management operations for ensuring Hihooi's fault recovery and proper scalability. 
This section explains these operations and discusses some enabling (future) work on automated backup 
and elasticity management.

\subsection{Backup and Fault Recovery}
\label{sec:scal:backup}

Hihooi needs to create backups to enable both recovery from failures and the efficient addition of new Extension DBs.
To avoid building its own complex backup and recovery procedures on top of the various database engines it manages, Hihooi uses the 
existing backup utilities offered by those engines. 
Even though the specific implementations differ, the overall process is the same: 
use the utilities to set a checkpoint in the database, associate with it the TSID of the last committed transaction, and create 
the backup (\textit{Seed DB}) in an incremental way.
In PostgreSQL, for example, this process is equivalent to creating a restore point and backing up the data up to that point.
In Oracle, on the other hand, it is equivalent to performing an online backup and replaying the redo log up to the checkpoint.
At the end, we get a read-consistent backup of the database at some point in time, while knowing which transactions it contains.

During and after the Seed DB creation, write transactions are modifying the Primary DB and are recorded into the Transactions Buffer.
As discussed in Section \ref{sec:architecture}, the Archiver periodically moves the transaction states that have been applied to all Extension 
DBs into the Archiver Buffer (in the form of text files) to keep the memory of the Transactions Buffer bounded.
In addition, after a Seed DB creation, the Archiver permanently removes from the Archiver Buffer all transactions that are already 
included in the Seed DB.
In case of a Primary DB failure, all write transactions from both buffers can be  replayed on the 
Seed DB node and the write workload can be transferred there (i.e., the Seed DB can also act as a hot standby node).
If the Archiver fails or the Archiver Buffer is lost, recreating the Seed DB resolves the issue and resets the Archiver Buffer since the Seed DB becomes up to date with the Primary DB.

The Transaction Manager (TM) builds the Transaction States (TStates) and pushes them on the Transactions Buffer as soon as transactions complete execution on the Primary DB (recall Section \ref{sec:repl:procedure}). 
However, TStates are deleted from the TM only after they are applied on all Extension DBs.
This greatly simplifies the recovery process of a potential Transactions Buffer failure as it guarantees that no TStates are lost.
If the Transactions Buffer fails, the Extension DBs will stop receiving any updates but will still be able to serve any read transactions that are consistent with their current database state.
Upon recovery of the Transactions Buffer, all collected TStates will be pushed from the TM to the Transactions Buffers and from there they will be applied on the Extension DBs in parallel.

A failure of the TM would sever all connections to the Primary DB, which would typically cause the abortion of active 
transactions \cite{middleware-si-sigmod05}. 
In the event of such failure, a new TM is launched on the backup node, which 
retrieves the previously active TStates from the Transactions Buffer and checks the Primary DB for their status.
If the committed transactions on the Primary DB are all reflected in the TStates on the Transactions Buffer, then 
(i) the new TM rebuilds its internal state and resumes operation,
(ii) the Extension DBs are notified of the change, and 
(iii) the clients are switched over to the new TM.
Otherwise, some transactions must have committed on the Primary DB but the TM failed before it could push the corresponding TStates to the Transactions Buffer.
In this case, a new Seed DB is incrementally created from the Primary DB and then applied to all Extension DBs, before the system resumes normal operations.
The same process is used in the event of a concurrent failure of the TM and the Transactions Buffer.

\subsection{Adding and Removing Extension DBs}
\label{sec:scal:adding}

The addition of a new Extension DB involves two steps: (i) the replication of the Seed DB on the new node, and (ii) 
the parallel re-execution of all transactions located on the Archiver Buffer using the procedure described in Section \ref{sec:repl:procedure}.
Afterwards, even though the new Extension DB might not be fully consistent with the Primary DB, it will register with the Transaction Manager 
and start serving consistent read requests, while applying the write transactions from the Transactions Buffer.
Hence, the addition of a new replica in Hihooi does not require a global synchronization barrier or the use of resources from other 
active replicas \cite{cecchet2008middleware}.

Extension DBs may be removed from the system for a variety of reasons such as maintenance operations, insufficient workload to justify 
their presence, and failures. Since Extension DBs only serve read transactions to the clients, no complicated failure mechanisms are needed 
from the client's perspective. The Transaction Manager is either notified or detects the removal of an Extension DB and simply 
re-routes the read transactions to other consistent Extension DBs. During the application of the write transactions from the Transactions Buffer, 
the Extractors log all completed transactions. When the node is re-added to the system, the write transactions are replayed from that point forward.

\subsection{Towards Replica Self-Management}
\label{sec:scal:elasticity}

The time required to start an Extension DB depends on the time needed to replicate the Seed DB (if the new Extension DB starts on a new node) plus the re-execution time 
of the write transactions on the Archiver Buffer. The former can be easily calculated since the process just entails bulk I/O transfers of known sizes.
The latter can also be computed because the execution time of each write transaction is recorded into the TState. 
Hence, Hihooi can accurately model and estimate the replica synchronization time. This model can guide the decision on how frequently to create new 
backups in order to provide bounded guarantees on the time needed to deploy a new Extension DB.

The ability to add and remove replicas without service interruption in addition to accurately modeling their cost are key steps towards autonomic 
middleware-based replicated databases. Recent work on database workload monitoring and characterization 
(e.g., \cite{workload-monitoring-sigmod11, workload-characterization-sigmod11}) could guide the development of elasticity policies that automatically 
decide when to add or remove nodes based on the actual workloads. Another interesting future direction would be integrating Hihooi with the 
cloud, which would extend the type or resources available for hosting the replicas. Finally, cloud technologies such as Virtual Machine 
migration or cloning could be used for creating the backups and launching new Extension DB nodes.

\section{Experimental Evaluation}
\label{sec:evaluation}

\begin{table}
	\centering
	\caption{Composition of TPC-C workload mixes}
	\label{tab:tpcc-workloads}
	\vspace{-1ex}
	\small
	{
	\begin{tabular}{l|cccc}
		\hline 
		Transaction & Read- & Read- & Balanced & Write- \\ 
		 & Only & Heavy &  & Heavy \\ 
		\hline 
		New-Order &  &  & 2\% & 7\% \\ 
		Payment &  & 3\% & 3\% & 5\% \\ 
		Order-Status & 50\% & 50\% & 85\% & 73\% \\ 
		Stock-Level & 50\% & 47\% & 10\% & 15\% \\ 
		\hline 
	\end{tabular} 
	}
	\vspace{-1ex}
\end{table}

\begin{figure*}[t!]
    \centering
	\subfloat[Read-Only]{%
	   \includegraphics[width=.175\textwidth]{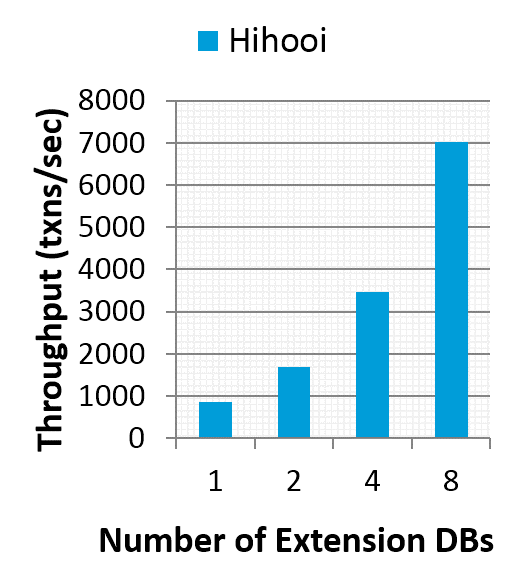}}\hfill
	\subfloat[Read-Heavy]{%
    	\includegraphics[width=.26\textwidth]{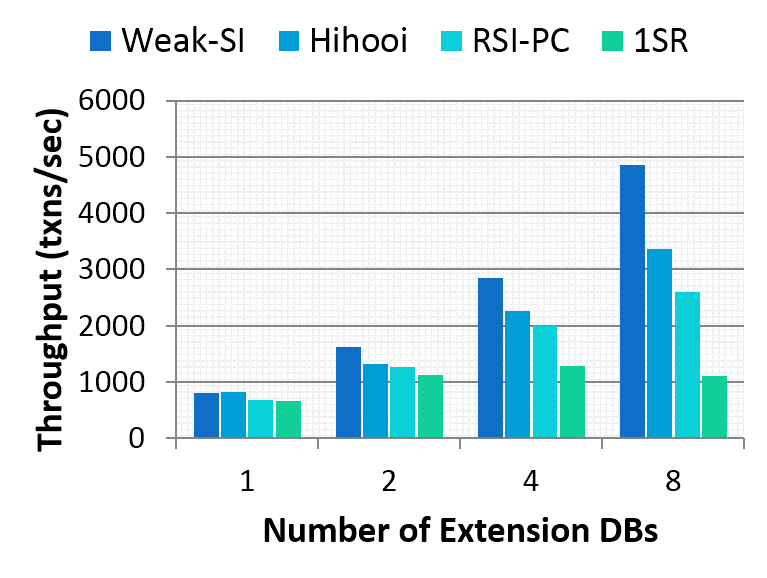}}\hfill
	\subfloat[Balanced]{%
    	\includegraphics[width=.26\textwidth]{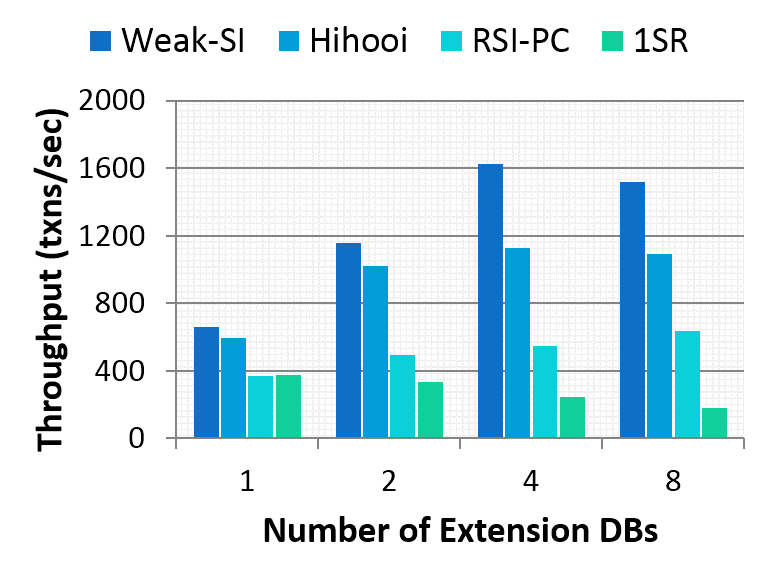}}\hfill
	\subfloat[Write-Heavy]{%
    	\includegraphics[width=.26\textwidth]{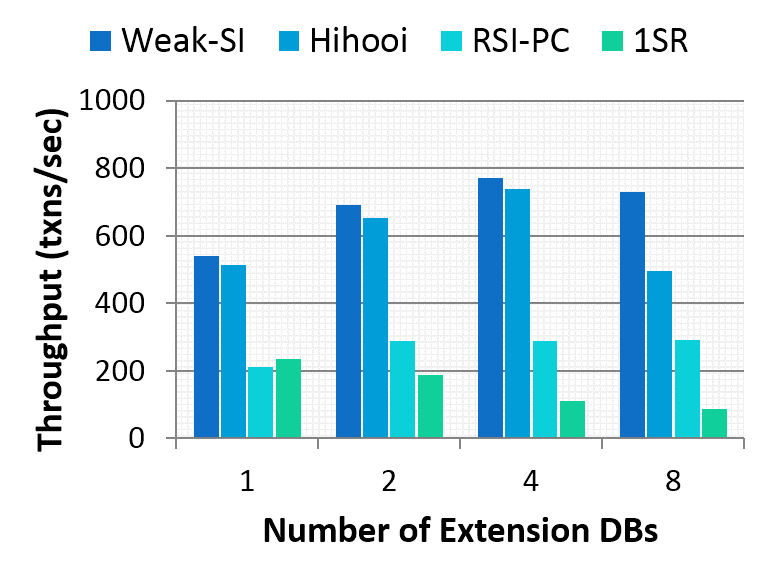}}
    \vspace{-1ex}
	\caption{OLTP workload scalability for TPC-C for different workload mixes and consistency levels}
  	\label{fig:test1_tpcc}
  	\vspace{-2.5ex}
\end{figure*}
	
\begin{figure*}[t!]
	\centering
	\subfloat[Read-Only]{%
		\includegraphics[width=.175\textwidth]{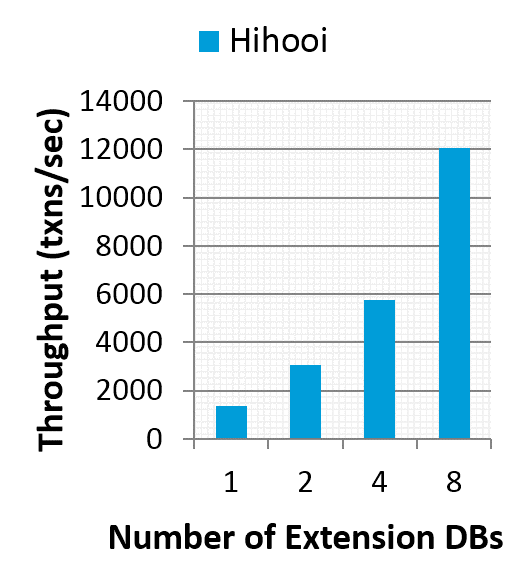}}\hfill
	\subfloat[Read-Heavy]{%
		\includegraphics[width=.26\textwidth]{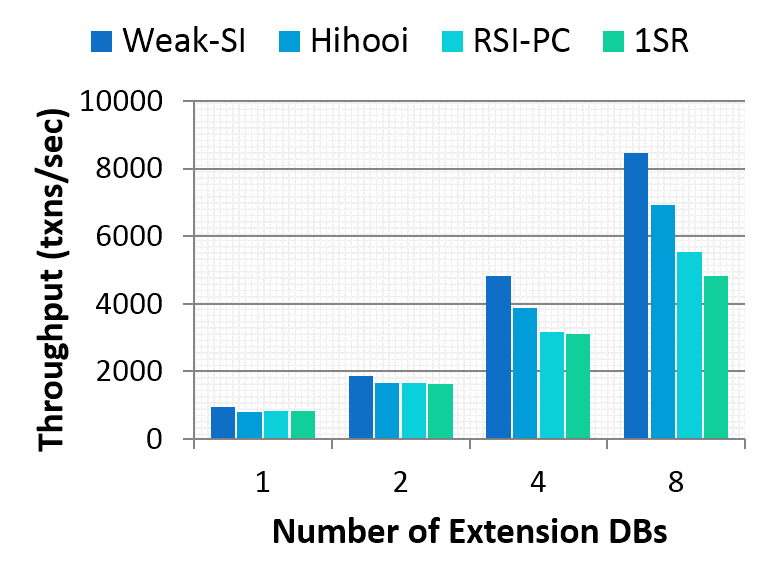}}\hfill
	\subfloat[Balanced]{%
		\includegraphics[width=.26\textwidth]{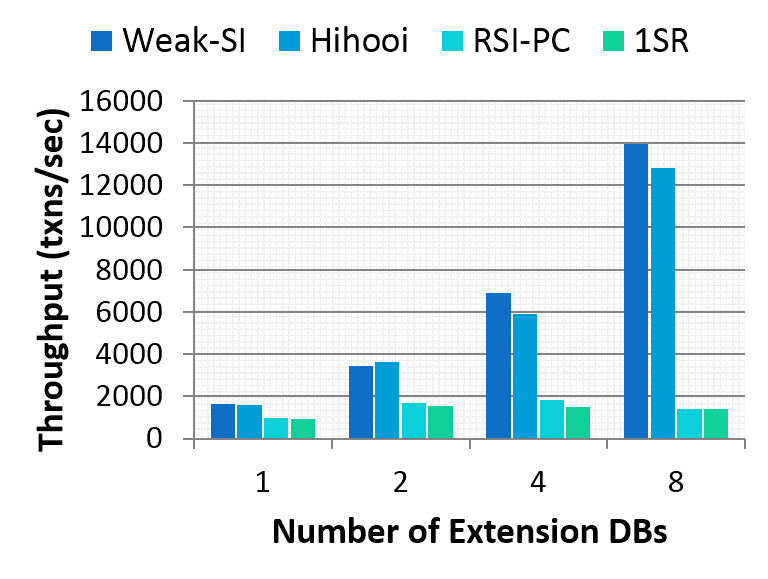}}\hfill
	\subfloat[Write-Heavy]{%
		\includegraphics[width=.26\textwidth]{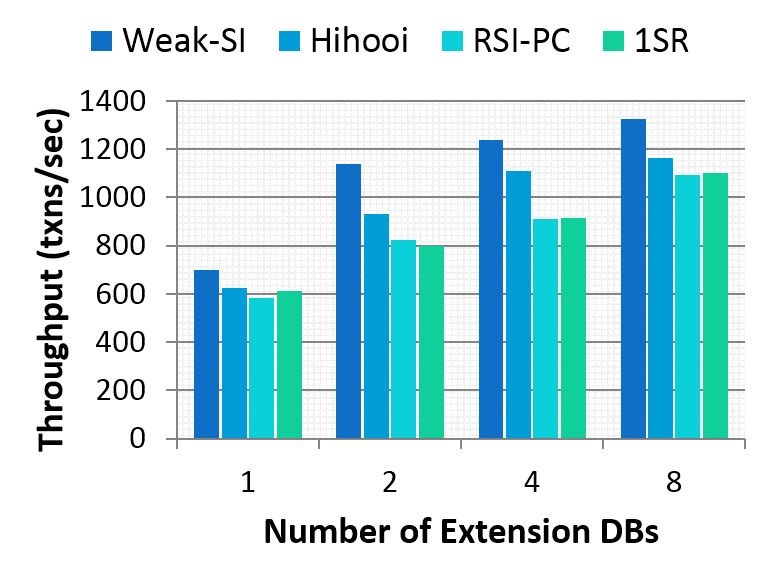}}
	\vspace{-1ex}
	\caption{OLTP workload scalability for YCSB for different workload mixes and consistency levels}
	\label{fig:test1_ycsb}
  	\vspace{-1ex}
\end{figure*}

The purpose of our evaluation is 
(1) to evaluate the system's performance and scalability under varying 
workload types and consistency levels,
(2) to study the effects of our fine-grained statement replication and routing algorithms, and
(3) to evaluate the key management and fault tolerance features of Hihooi.
All experiments were run on a 13-node cluster running CentOS Linux 7.2 with 1 Primary DB, 1 Seed DB, 8 Extension DBs, and 3 client nodes (with up to 16 clients each).
The Transaction Manager is running on the Primary DB node, its backup and Archiver on the Seed DB node, and the Transactions Buffer on an active Extension DB node.
The primary node has an 8-core, 3.2GHz CPU, 64GB RAM, and 2.1TB HDD storage.
The rest nodes have an 8-core, 2.4GHz CPU, 24GB RAM, and 1.5TB HDD storage.

For our evaluation, we used three well-known benchmarks, each employing a different kind of workload:
(i) \textbf{TPC-C} \cite{tpcc}, the industry standard for OLTP workloads, containing complex and write-intensive transactions;
(ii) \textbf{YCSB} (Yahoo Cloud Serving Benchmark) \cite{ycsb-socc10}, a collection of web-based micro-benchmarks that represent data 
management applications whose workload is simple but requires high scalability; and
(iii) \textbf{CHB} (CH-benCHmark) \cite{chbenchmark}, a workload combining OLTP from TPC-C and OLAP from TPC-H \cite{tpch}.

The TPC-C database (also used by CHB) was populated with 500 warehouses for a total size of 50GB.
For the YCSB database, we used a scalefactor of 50000, resulting in 56GB of data.
The databases were fully replicated to the Extension DBs.
We used PostgreSQL version 9.5.3 in all nodes. 
The results presented, unless noted otherwise, 
are from 10 minute trials, preceded by 2 minutes of warm up.
OLTP-Bench~\cite{oltpbench} was used to populate and run the tests for all benchmarks.
The transactions load was injected using 6 clients per 1 Extension DB that continuously issued transactions.

\subsection{OLTP Workload Scalability}
\label{sec:eval:oltpscaleup}

This section studies the effectiveness and efficiency of Hihooi in scaling an OLTP workload by measuring its throughput and latency as we increase the number of Extension DBs. 
The comparison is done along two dimensions: 
(i) for different read/write workload mixes (i.e., Read-Only, Read-Heavy, Balanced, and Write-Heavy) and
(ii) for different consistency levels (i.e., Weak-SI, Hihooi, RSI-PC, and 1SR; recall Section \ref{sec:cc:levels}).
Weak-SI is used to show the upper limit of performance that any system with consistency guarantees could achieve.
RSI-PC is used by Ganymed, a similar middleware system that does not offer the type of replication and routing algorithms that Hihooi boasts, while 1SR (used by C-JDBC) shows the effect of synchronous replication.

For TPC-C, the Read-Only, Read-Heavy, Balanced, and Write-Heavy workload mixes were set up as 100\%, 95\%, 85\%, and 70\% of read statements, respectively,
and were generated via mixing the TPC-C transactions as shown in Table \ref{tab:tpcc-workloads}.
Figure \ref{fig:test1_tpcc} shows the throughput rates in committed transactions per second for our workload mixes and consistency levels. 
The Read-Only workload scales linearly as the number of replicas increases; that is, the throughput doubles each time the number of Extension DBs doubles.
As no writes are performed, there is no difference between the 4 consistency levels.
The trend is similar for the Read-Heavy workload, with the exception of 1SR after 4 or more replicas are used.
This is expected since the system has to wait for more replicas to apply all modifications before being able to serve any subsequent reads.
As the percentage of writes increases in the workload, scalability naturally suffers for all consistency levels, since 
all writes are executed on the Primary DB and more reads have to wait for a consistent replica.
Nonetheless, \textit{Hihooi is always able to offer comparable performance to Weak-SI} and up to 2.6x and 6.7x higher throughput compared to RSI-PC and 1SR, respectively.

\begin{figure*}[t!]
	\centering
	\subfloat[Read-Heavy]{%
		\includegraphics[width=.31\textwidth]{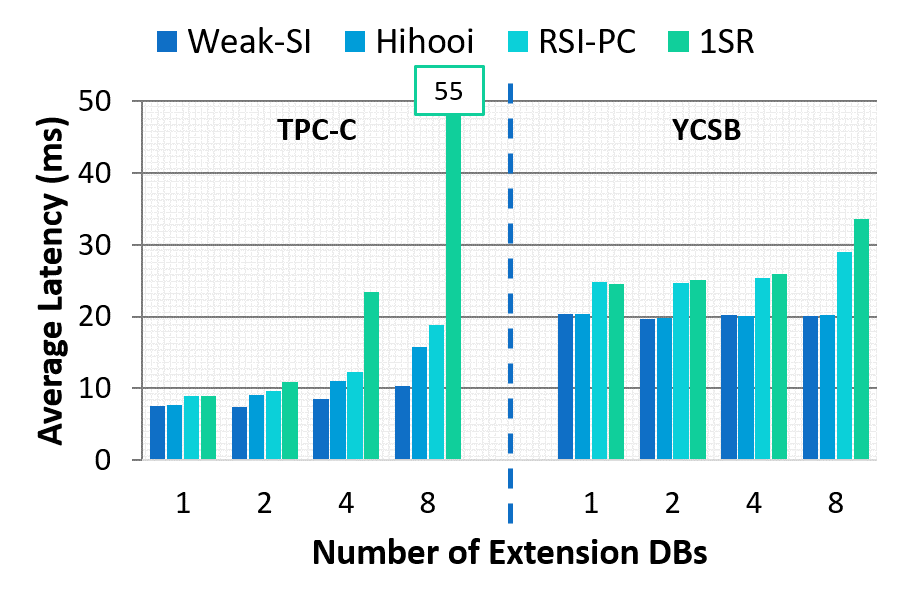}}\hfill
	\subfloat[Balanced]{%
		\includegraphics[width=.31\textwidth]{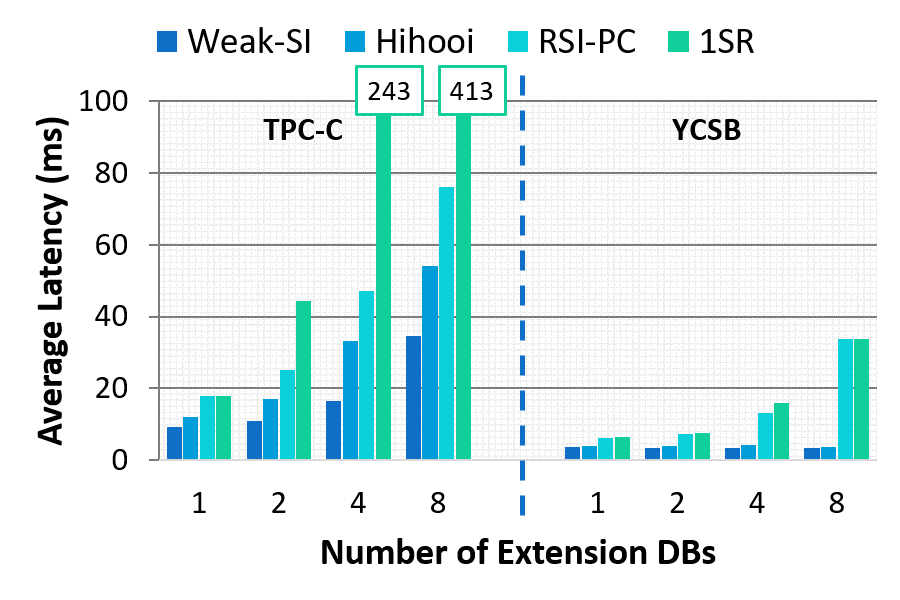}}\hfill
	\subfloat[Write-Heavy]{%
		\includegraphics[width=.31\textwidth]{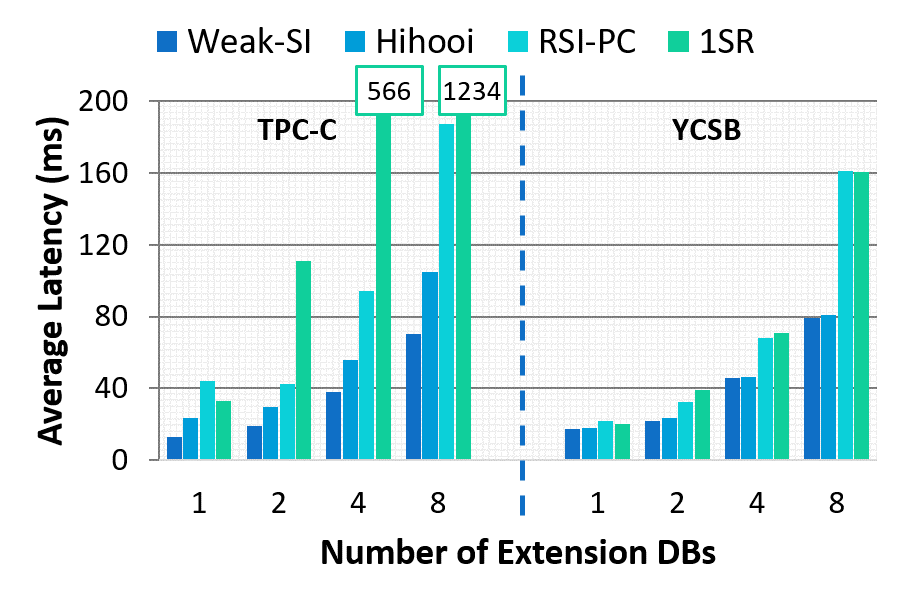}}
	\vspace{-1ex}
	\caption{Average latency for TPC-C and YCSB for different workload mixes and consistency levels}
	\label{fig:test1_latency}
	\vspace{-1.7ex}
\end{figure*}

Figure \ref{fig:test1_ycsb} shows the throughput rates for four workload mixes and our consistency levels for YCSB.
All YCSB workloads follow a Zipfian distribution (theta=1) and contain unmodified queries.
Similar to TPC-C, the Read-Only YCSB workload exhibits almost perfect linear scalability.
The Read-Heavy workload consists of 5\% inserts and 95\% range scan queries, per \cite{ycsb-socc10},
while the Balanced workload consists of 85\% single-row reads and 15\% inserts.
Both Weak-SI and Hihooi are still able to achieve near linear scalability, while RSI-PC and 1SR do not scale at all for the Balanced workload.
Based on our observations, inserts in YCSB are 2-3x faster than reads.
In Ganymed, this results in delays in the execution of reads, which need to wait for all fast inserts to propagate to at least one replica.
The use of row R/W sets by Hihooi's routing algorithms excels in this test as it enables the system to route a read transaction 
$T_r$ to a replica that might not be fully consistent with the Primary DB but is consistent for $T_r$.
Finally, the Write-Heavy workload consists of 50\% reads and 50\% updates, per \cite{ycsb-socc10}.
As this is a more demanding workload, both throughput and scalability suffer.
Nevertheless, \textit{Hihooi still performs considerably better (up to 1.42x) compared to RSI-PC and 1SR} for the same 
reasons.

\begin{figure*}[t]
	\begin{minipage}[t]{0.298\linewidth}
		\vspace{0pt}
		\centering
		\includegraphics[width=.98\textwidth]{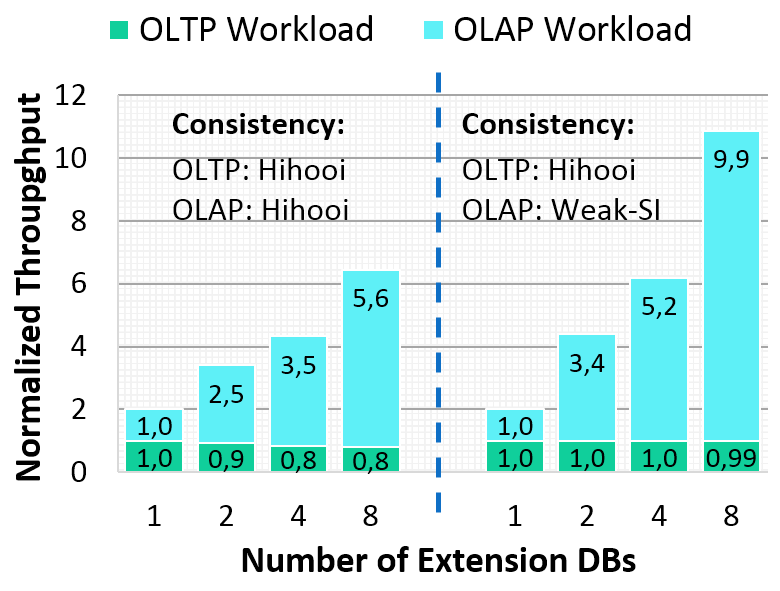}
		\vspace{-1ex}
		\captionsetup{justification=raggedright}
		\caption{Mixed OLTP-OLAP workload scalability for CHB}
		\label{fig:test2_all}
	\end{minipage}
	\hfill
	\begin{minipage}[t]{0.33\linewidth}
		\vspace{0pt}
		\centering
		\includegraphics[width=.98\textwidth]{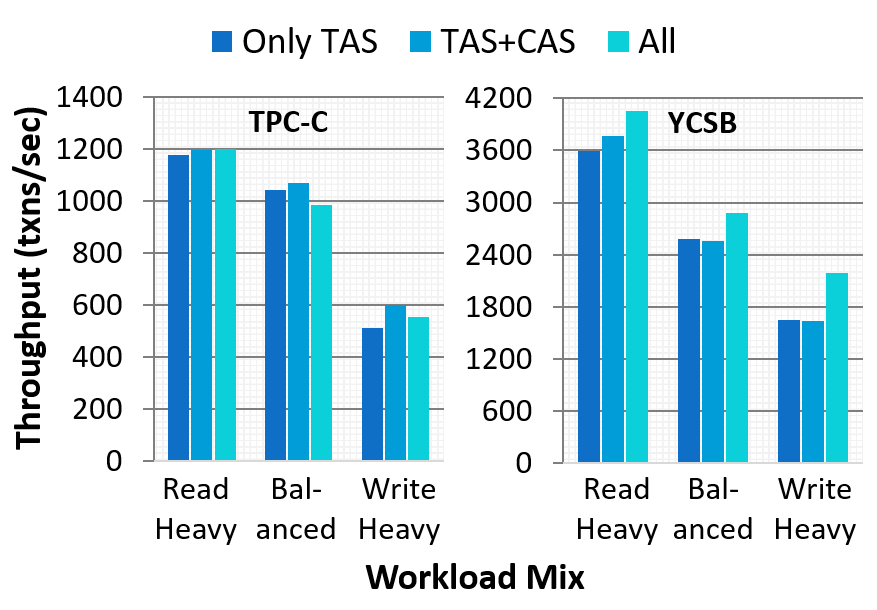}
		\vspace{-1.3ex}
		\captionsetup{justification=raggedright}
		\caption{Effect of using the TAS, CAS, ~~~and RAS classes}
		\label{fig:test3_all}
	\end{minipage}
	\hfill
	\begin{minipage}[t]{0.355\textwidth}
		\vspace{0pt}
		\centering
		\includegraphics[width=.98\textwidth]{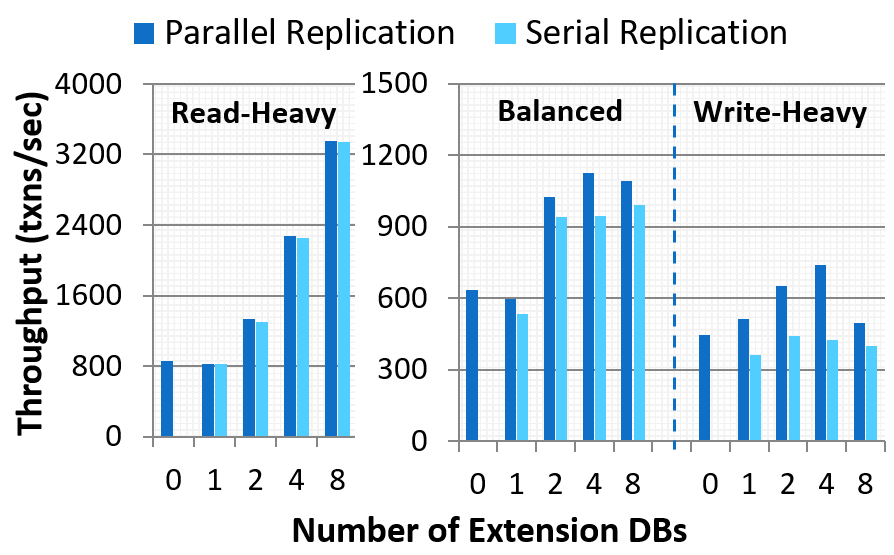}
		\vspace{-1ex}
		\caption{Effect of Hihooi's parallel replication algorithm on TPC-C}
		\label{fig:test5_tpcc}
	\end{minipage}
	\vspace{-0.5ex}
\end{figure*}

\begin{figure*}[t]
	\begin{minipage}[t]{0.34\linewidth}
		\vspace{0pt}
		\centering
		\includegraphics[width=.98\textwidth]{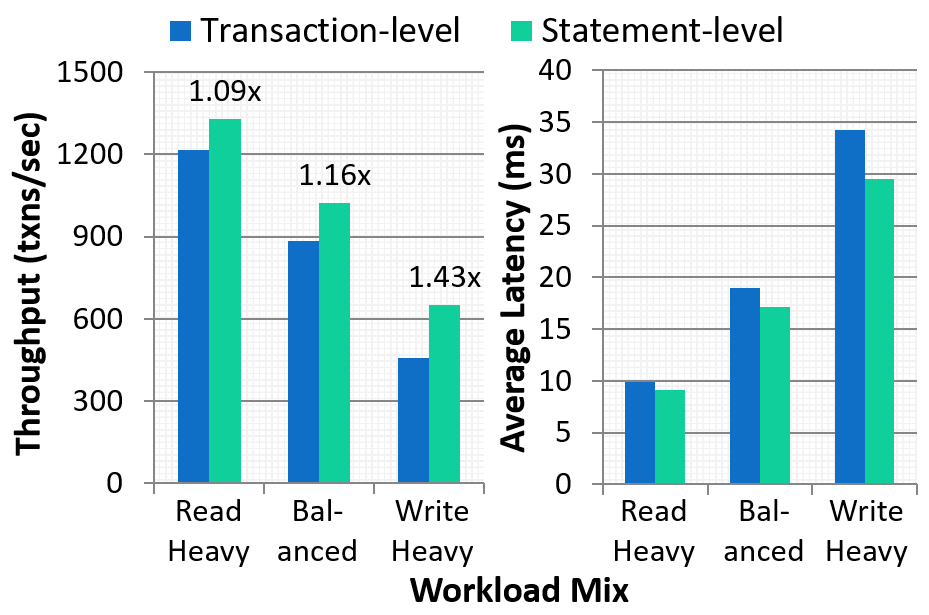}
		\vspace{-0.8ex}
		\captionsetup{justification=raggedright}
		\caption{Effect of statement-level load balancing on TPC-C}
		\label{fig:test4_tpcc}
	\end{minipage}
	\hfill
	\begin{minipage}[t]{0.31\linewidth}
		\vspace{0pt}
		\centering
		\includegraphics[width=.98\textwidth]{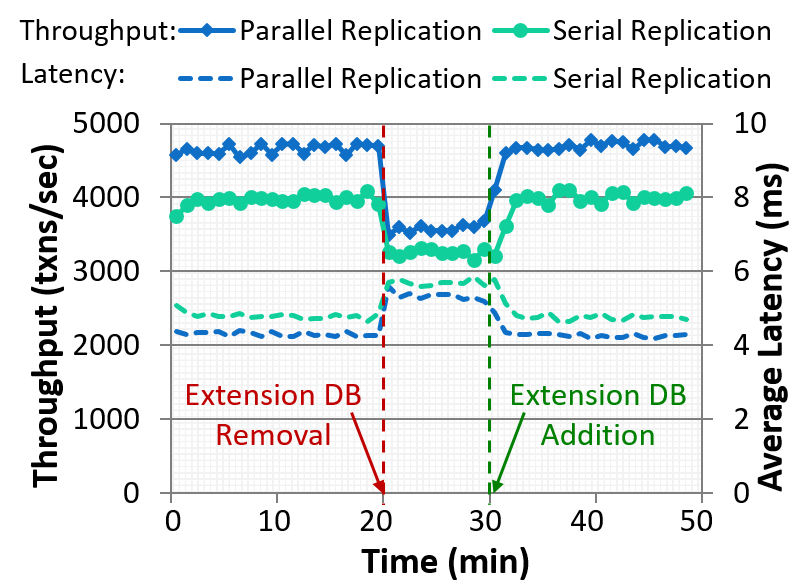}
		\vspace{-1ex}
		\captionsetup{justification=raggedright}
		\caption{YCSB throughput after removing \& adding 1 Extension DB}
		\label{fig:test6_ycsb}
	\end{minipage}
	\hfill
	\begin{minipage}[t]{0.31\textwidth}
		\vspace{0pt}
		\centering
		\includegraphics[width=.98\textwidth]{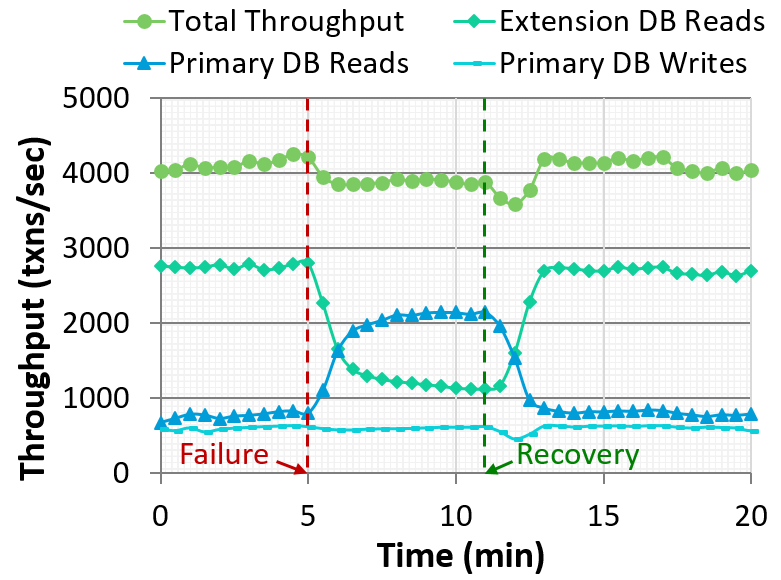}
		\vspace{-1ex}
		\caption{YCSB throughput during Transactions Buffer failure}
		\label{fig:trans-buffer-failure}
	\end{minipage}
	\vspace{-2ex}
\end{figure*}

Figure \ref{fig:test1_latency} shows the average latency of transactions across different workloads mixes for TPC-C and YCSB.
For the Read-Heavy workloads, there is very little to no increase in latencies as the number of replicas increases due to the efficient load balancing of read queries to the replicas.
However, as more writes are introduced in the Balanced and Write-Heavy workloads, adding more replicas increases the average latencies (even for Weak-SI) as a bigger percentage of the workload is sent to the Primary DB.
Focusing on Hihooi, we observe only a small increase in latency as the number of replicas increase, indicating the low overhead added due to replication.
Conversely, both RSI-PC and 1SR cause increasingly larger latencies for all workloads due to waiting reads (as explained above). 
Once again, \textit{Hihooi is able to offer latencies comparable to Weak-SI} in almost all cases due to its fine-grained transaction routing capabilities.

\subsection{OLTP-OLAP Workload Scalability}
\label{sec:eval:oltpolapscaleup}

The execution of OLAP queries on transactional databases has long been a motivating scenario for database 
replication \cite{cecchet2008middleware, middleware-si-sigmod05}. 
In this section, we evaluate the OLAP workload scalability provided by 
Hihooi, while studying its effects on an OLTP workload. 
For these tests, an OLTP client node executes the CHB transactional workload, while two OLAP client nodes submit the CHB analytical queries, all using the default Hihooi consistency level (GSSI).
The general trend, as shown in Figure \ref{fig:test2_all}, is that the OLAP workload scales sub-linearly, while the OLTP one exhibits a small negative impact that worsens as the number of replicas increases (7-20\%) because
more OLAP queries are forced to execute on the Primary DB (due to read-write conflicts).
However, since OLAP workloads do not typically require strong consistency, we repeated the experiment using Weak-SI for the OLAP workload and GSSI for the OLTP one.
The new results (see Figure \ref{fig:test2_all}) reveal \textit{linear scalabilty for the OLAP workload with almost no overhead for the OLTP one}, and highlight the great benefits offered by Hihooi in this setting.

\subsection{Effect of Affecting Classes}
\label{sec:eval:classes}

\begin{table}
	\centering
	\caption{Percentage (\%) of TAS, CAS, and RAS statements}
	\label{tab:affecting-classes}
	\vspace{-1ex}
	\small
	{\setlength{\tabcolsep}{0.6em}
	\begin{tabular}{l l | c c c}
		\hline 
		Benchmark & Workload & \multicolumn{3}{c}{Affecting Class} \\ 
		&  & TAS & CAS & RAS \\ 
		\hline 
		TPC-C & Read-Only & 0 & 10 & 90 \\ 
		& Read-Heavy & 4 & 10 & 86 \\ 
		& Balanced & 23 & 14 & 63 \\ 
		& Write-Heavy & 40 & 12 & 48 \\ 
		\hline 
		YCSB & Read-Only & 0 & 0 & 100 \\ 
		& Read-Heavy & 95 & 0 & 5 \\ 
		& Balanced & 15 & 0 & 85 \\ 
		& Write-Heavy & 50 & 0 & 50 \\ 
		\hline 
		CHB & OLTP & 56 & 13 & 31 \\ 
		& OLAP & 82 & 18 & 0 \\ 
		\hline 
	\end{tabular} 
	}
\vspace{-1.5ex}
\end{table}


Query statements can modify or access data at different levels of granularity, namely at the table, column, or row level, captured by our 
definitions of TAS, CAS, and RAS classes, respectively (recall Section \ref{sec:repl:rwsets}).
Table \ref{tab:affecting-classes} lists the percentage of TAS, CAS, and RAS statements for each workload mix of each benchmark.
All TPC-C workloads contains a mix of all three types of statements, while the percentage of TAS increases as the workload becomes more write-heavy.
YCSB workloads, on the other hand, contain a significant fraction of RAS and no CAS statements.

This section studies the effect of letting Hihooi use these increasing levels of granularity by configuring it to use: 
(i) only TAS; (ii) TAS and CAS; and (iii) all classes.
We executed both the TPC-C and YCSB workloads with our different mixes on Hihooi using two Extension DBs.
The results presented in Figure \ref{fig:test3_all}	reveal that \textit{the benefits from utilizing the CAS and RAS classes depend both on the read-write mix and the workload itself}. 
In particular, the benefits for read-heavy workloads are relatively small because the replicas are almost always consistent with the Primary DB and the reads are typically load-balanced regardless of their affecting class.
As the write portion of a workload increases, there are more opportunities for Hihooi to route read statements that access some columns or rows of a table, 
even though some other columns or rows of that same table have been modified.
This is more evident with YCSB, whose Balanced and Write-Heavy workloads contain a significant fraction of RAS statements.
Hence, when Hihooi is able to exploit RAS, the overall throughput is increased up to 32\% in our experiments.
TPC-C on the other hand, uses more complex transactions 
that effect tables both at the column or row level, 
leading to small benefits from using CAS or RAS (less than 22\%).
The effects on average latency follow the same trends as the effects on throughput shown in Figure \ref{fig:test3_all} and are not shown due to space constraints.
Finally, the memory overhead from the hash indexes is very low as the maximum one measured in all experiments was less than 300 KB.

\vspace{-1ex}
\subsection{Parallel Replication Algorithm}
\label{sec:eval:parallel}

This section delves into the performance implications of the parallel replication algorithm (recall Section \ref{sec:repl:procedure}) compared to the common approach that executes the write transactions serially on the replicas.
The two approaches have no to little impact on the throughput of the Read-Only \& Read-Heavy TPC-C workloads (see Figure \ref{fig:test5_tpcc}) since very few writes are applied to the replicas.
Note that TPC-C contains 1 TAS and 11 RAS write statements, which are amenable to parallelism.
The actual degree of parallelism (\textit{dp}) when applying the writes depends on the submission order of the writes as well as the portion of the data they apply to.
As an indicative example, the average \textit{dp} for the Balanced workload with 4 Extension DBs was 4.
As the percentage of writes increases for the Balanced and Write-Heavy workloads, \textit{the parallel algorithm has a profound effect on the throughput} (up to 1.7x higher compared to the serial version) because it
enables the Extension DBs to reach consistency quicker and, hence, be available to serve more reads.
There is a drop in performance for the write-heavy workload on 8 Extension DBs because the writes generated concurrently by the 48 clients overload the Primary DB.
At that point, even the Weak-SI case experienced a performance drop (see Figure \ref{fig:test1_tpcc}(d)).

The 0-Extension DBs setting in Figure \ref{fig:test5_tpcc} corresponds to processing the workload on a single node without replication.
The difference between having 0 and 1 Extension DBs reveals the overhead incurred by Hihooi from intersecting all transactions, which was typically low ($<$4\%) and no more than 9\% across all experiments (not shown due to space constraints).
It is interesting to note that for heavy-write workloads, Hihooi is very effective in separating the execution of writes and reads on the Primary and Extension DBs, respectively, leading to an aggregated higher throughput.

\vspace{-1ex}
\subsection{Statement-level Load Balancing}
\label{sec:eval:stmt}

One of the most novel aspects of Hihooi is its ability to route individual read statements to consistent replicas, 
even within multi-statement write transactions. 
This section evaluates the effect of statement-level versus 
(the typical) transaction-level load balancing, which always routes all statements from a write transaction to the Primary DB.
Figure \ref{fig:test4_tpcc} shows the throughput and average latency for our 3 TPC-C workloads
executed on Hihooi running with two Extension DBs using either transaction- or statement-level load balancing.
\textit{As the percent of writes increases, so does the benefit of statement-level load balancing}, 
leading up to 1.43x better throughput and 14\% lower latency compared to transaction-level load balancing.
These benefits are attributed to the extra read statements that are diverted to the Extension DBs.
Specifically, in the Write-Heavy workload, the transaction-level algorithm routes 29\% of the reads to the Primary DB either because there are no consistent Extension DBs or the reads are part of multi-statement write transactions.
On the contrary, the statement-level algorithm routes only 15\% of reads to the Primary DB, while the remaining are load balanced to the Extension DBs.
Overall, with more multi-statement write transactions, Hihooi has more opportunities to divert the included reads to Extension DBs, increasing parallelism and, therefore, throughput.

\vspace{-0.5ex}
\subsection{Adding and Removing Extension DBs}
\label{sec:eval:addremove}

Next, we explore the scenario of adding and removing an Extension DB at run time.
We started running the Balanced YCSB workload on Hihooi with 18 Clients and 3 Extension DBs.
After 20 minutes, we removed 1 Extension DB to simulate a failure or planned maintenance operation.
The read transactions executing on the Extension DB failed at that point but the Transaction Manager automatically rerouted them to other replicas.
Hence, Hihooi continued serving the workload without any issues, 
albeit with a 24\% lower throughput and higher average latency, as shown in Figure \ref{fig:test6_ycsb}.
After 10 minutes, we restored the Extension DB and observed the throughput rate return to its normal level quickly.
\textit{The Extension DB was able to serve its first read just 64 seconds after restoration} due to our fine-grained routing algorithm, while it was able to apply all changes it missed during the outage in 82 seconds.
In total, it missed 321786 write transactions, while the memory size of the Transactions Buffer grew to only 384MB.
We repeated the above procedure using the serial replication approach and observed a lower throughput and higher average latency during the entire experiment, while it took the Extension DB 121 seconds to catch up; highlighting once again the benefits of our parallel replication algorithm.

\vspace{-0.5ex}
\subsection{Transactions Buffer Failure and Recovery}
\label{sec:eval:tbuffer-failure}

In this section, we investigate the behavior of Hihooi during the failure and recovery of the Transactions Buffer.
We started running the Balanced YCSB workload on Hihooi with 24 Clients and 4 Extension DBs.
After 5 minutes, we induced a failure on the Transactions Buffer, which caused the Extension DBs to stop receiving any updates.
However, Hihooi kept serving the incoming workload without any query failures.
As the Extension DBs kept falling behind, the amount of read transactions executing on them decreased (since the YCSB workload is skewed to favor recent items), while many read transactions were routed towards the Primary DB, as shown in Figure \ref{fig:trans-buffer-failure}.
The write throughput was unaffected by the failure due to the asynchronous nature of the replication procedure.
Overall, the total throughput experienced a small slowdown of only 6.2\%.
After 6 minutes, we recovered the Transactions buffer.
At that point, all Transaction States accumulated at the Transaction Manager (207220 in total, 247MB in size) were pushed on the Transactions Buffer and the Extension DBs started applying them in parallel.
The overhead caused by the recovery process led to a small, 5.4\% decrease in the overall throughput of the workload, which lasted for only 97 seconds until it returned back to its pre-failure level.
These results show that Hihooi is able to gracefully handle a Transactions Buffer failure.

\subsection{Comparison with PostgreSQL Replication}
\label{sec:eval:pgpool}

\begin{figure}[t!]
	\centering
	\includegraphics[width=.36\textwidth]{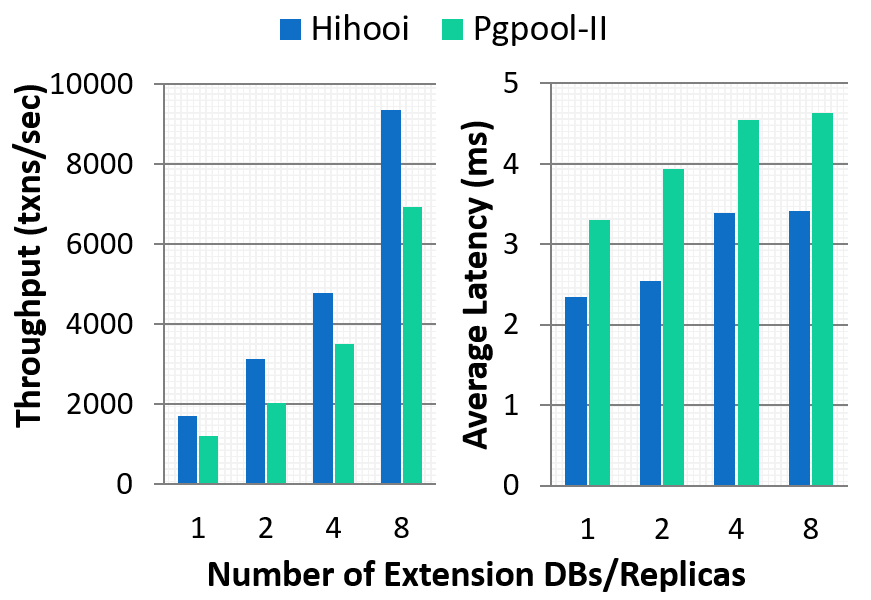}
	\caption{Pgpool-II Vs Hihooi for YCSB Balanced workload}
	\label{fig:pgpool-vs-hihooi-ycsb}
	\vspace{-2pt}
\end{figure}

PostgreSQL supports master-slave replication, where the master database server executes both read and write transactions and the slave (``hot standby'') replicas execute only read queries.
PostgreSQL replicates database modifications via streaming WAL records from the master to the replicas, i.e., it employs \textit{row-based replication}.
This replication is asynchronous by default so the data on the standby is eventually consistent with the primary.
On top of a PostgreSQL cluster, Pgpool-II \cite{postgresql-book15} is used to provide connection pooling and load balancing of read queries to the replicas.
We have setup Pgpool-II with PostgreSQL replication on our local cluster and compared its performance against Hihooi.

Figure \ref{fig:pgpool-vs-hihooi-ycsb} shows the throughput and average latency of the YCSB Balanced workload when executed on Pgpool-II and Hihooi as the number of replicas increases.
Our results show that both systems can scale throughput with more replicas in a similar fashion, while having a small negative impact on the average latency.
Nonetheless, Hihooi is able to offer 35-55\% higher throughput and 26-35\% lower latency compared to Pgpool-II across all experiments.
Hihooi's better performance is attributed to
(i) its parallel replication algorithm (as PostgreSQL applies the WAL records serially), and
(ii) its fine-grained routing algorithm (as Pgpool-II does not load-balance multi-statement write transactions).

\section{Related Work}
\label{sec:related}

Database replication comes in two forms:
(i) \textit{master-slave}, where one primary copy handles all writes and the other replicas process only reads \cite{ganymed-middleware2004, middle-r-2005, byz-faults-sosp07}; and 
(ii) \textit{multi-master}, where all replicas serve both 
reads and writes \cite{c-jdbc, postgres-r, tashkent-2006}.
Each form can be implemented either inside the database kernel or outside in a middleware layer.
While the former approach provides opportunities for various optimizations and a tight coupling of concurrency and 
replica control, it is heavily invasive and database-engine specific \cite{kemme2010database}.
The middleware approach, also employed by Hihooi, leads to a seamless separation of concerns, supports unmodified database systems and applications, 
and can enable heterogeneous environments.


Postgres-R \cite{postgres-r} was one of the first multi-master replication systems to use group communication primitives with 
strong ordering to enable scalability and 1-copy-serializabi-lity, while a later version offered snapshot isolation (SI) \cite{postgres-r-si-icde05}.
Middle-R \cite{middle-r-2005} was the middleware extension of Postgres-R that moved group communication outside the database engine but 
still required database modifications for extracting and applying tuple-based updates. Other similar systems that rely on group communication 
primitives and offer SI are Tashkent \cite{tashkent-2006} and SI-Rep \cite{middleware-si-sigmod05}.
C-JDBC \cite{c-jdbc} is also a multi-master middleware system but does not require database modifications as it uses JDBC drivers like Hihooi.
The system offers consistency guarantees through table-level locking at the middleware level.

DBFarm \cite{dbfarm-middleware06} builds upon ideas from Middle-R 
(and thus requires database engine modifications) but 
offers a master-slave middleware system. 
As such, a read transaction is delegated to some replica 
but it is blocked until that replica is consistent with the primary. 
\cite{distr-versioning, icdcs02-Jimenez} present middleware solutions that require a predeclaration of the access pattern of all transactions to enable efficient scheduling.
In \cite{byz-faults-sosp07}, the middleware will first execute a write transaction on the primary replica, extract lock-based concurrency information, and use that to enforce a transaction scheduling to the replicas, which prevents conflicting schedules.
Unlike Hihooi, \cite{byz-faults-sosp07} requires the underlying databases to use strict two-phase locking and cannot handle snapshot isolation, which is now widely used.
Ganymed \cite{ganymed-middleware2004} is a similar middleware 
system that instead blocks a read transaction at the middleware layer until at least one replica becomes consistent.
On the contrary, Hihooi never blocks any read transactions. 
Rather, it uses the transaction read/write sets to find the replicas,
including the Primary DB, that are consistent for each read transaction to run on.
In doing so, Hihooi is the first replication-based middleware to offer such fine-grained statement-based routing, even within multi-statement write transactions.
Pgpool-II \cite{postgresql-book15} is another PostgreSQL-specific replication middleware solution that ships and applies WAL entries to the replicas.
Pgpool-II, similar to DBFarm and Ganymed, apply all database modifications serially at the replicas, as opposed to Hihooi that applies them in parallel.

Another way in which Hihooi differs from the state of the art is its new architecture that uses an in-memory distributed storage system for 
statement replication, rather than relying on command logging propagation or complex group communication protocols \cite{middleware-si-sigmod05, cecchet2008middleware}. 
The Transactions Buffer acts as a highly available 
propagation medium for all database modifications that need to be applied asynchronously to active 
replicas, improving network load distribution and simplifying recovery procedures.
Amazon Aurora \cite{verbitski2017amazon} has a different architecture that decouples compute from storage while employing primary copy replication to achieve read scale-out.
Aurora uses physical replication, where the redo log records are replayed in the replicas, allowing them to be physically identical to the primary. 
Such an approach, however, cannot be used to scale existing single-node databases (unlike Hihooi).

Other systems such as Hyder \cite{bernstein2011hyder,bernstein2015optimizing} and Tango \cite{balakrishnan2013tango} provide the abstraction of a replicated in-memory data structure backed by a shared log, and leverage the shared log to enable fast transactions across different objects.
\cite{wang2017query} and \cite{qin2017scalable} provide log shipping from a primary copy. 
The former uses synchronous writes so it avoids concurrency issues from reading from replicas, but it relies on the presence of InfiniBand and NVRAM to be efficient.
The latter replays logs at the level of records but the approach only targets the scenario of primary-backup replication with a single backup instead of multiple replicas.
KuaFu \cite{kuafu} is a primary-backup, row-based replication system that offers concurrent log replay by constructing and utilizing a graph to track write-write dependencies in the log; unlike Hihooi that relies solely on TSIDs and read/write sets.
To allow read operations to be served on backups, KuaFu introduces barriers every $N$ transactions to create snapshots that are consistent with some past states on the primary, unlike Hihooi that never uses barriers.

Commercial clustering solutions such as Oracle RAC \cite{oraclerac} and IBM DB2 pureScale \cite{ibmpurescale} rely on the use of specialized 
hardware and network-attached storage to work. Hence, unlike our approach, the system cannot easily be installed on a set of commodity servers.
Finally, other database replication products such as Oracle Golden Gate \cite{gupta2013oracle} exist, but only offer weak consistent properties and 
are meant to be used for off-line reporting or disaster recovery plans.

Data partitioning is another popular scale-out approach that partitions and distributes data across cluster nodes \cite{scalable-sql-sigmodrec11, calvin-sigmod12}.
Such approaches are amenable to dynamic scaling via migrating data to existing or new nodes in order to diminish performance issues due to skew or heavy loads.
Accordion \cite{accordion-vldb14} migrates data at a coarse predefined granularity, whereas E-Store \cite{estore-vldb14} and Clay \cite{clay-vldb16} work at a finer tuple-level granularity.
The aforementioned approaches perform data migrations after detecting performance issues, whereas P-Store \cite{pstore-sigmod18}, another elastic OLTP DBMS, focuses on workload prediction and proactive migration.
One of the key scalability hurtles of data partitioning approaches are transactions spanning multiple partitions as they require locking or other specialized protocols; a non-existent issue for Hihooi as all transactions have access to the full database.
Finally, the issue of dynamic scaling is orthogonal to our approach and something we plan to work on in the near future.

\section{Conclusions}
\label{sec:conclusions}

\hyphenation{data-bases}

As a replication-based middleware system, Hihooi is able to provide workload scalability to existing databases without sacrificing consistency.
The parallel replication algorithm 
allows the Extension DBs to reach consistency quicker, while the routing algorithm avoids any delays by routing read statements to consistent replicas.
We believe Hihooi can jump start interesting research towards automated elasticity as well as the creation of new cloud-based offerings.


\balance

\bibliographystyle{IEEEtran} 
\bibliography{paper}


\vskip -2\baselineskip plus -1fil
\begin{IEEEbiography}[{\includegraphics[width=1in,height=1.25in,clip,keepaspectratio]{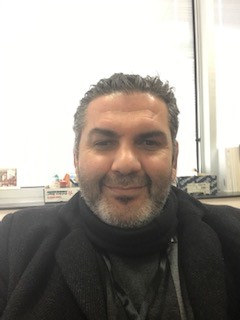}}]{Michael A. Georgiou}
	is a PhD Candidate at the Cyprus University of Technology and works at the National Bank of Greece (Cyprus) as a Senior Oracle Database Administrator.
	He is an Oracle Certified Professional, Oracle RAC Expert, and Red-Hat Certified Engineer. 
	His research interests are in parallel and distributed databases, cloud database technologies, and dynamic workload management.
\end{IEEEbiography}

\vskip -2\baselineskip plus -1fil
\begin{IEEEbiography}[{\includegraphics[width=1in,height=1.25in,clip,keepaspectratio]{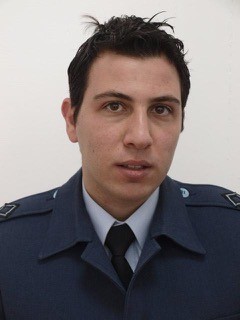}}]{Aristodemos Paphitis}
	is a PhD Candidate at the Cyprus University of Technology.
	He holds a MSc from University of Cyprus on Advanced Technologies in Computer Science
	and BSc in Electronics and Telecommunications Engineering from the Hellenic Airforce Academy.  
	His research interests lie in the fields of wireless networks,
	transactional workload scalability, 
	blockchain technologies, and repeatability engineering.
\end{IEEEbiography}

\vskip -2\baselineskip plus -1fil
\begin{IEEEbiography}[{\includegraphics[width=1in,height=1.25in,clip,keepaspectratio]{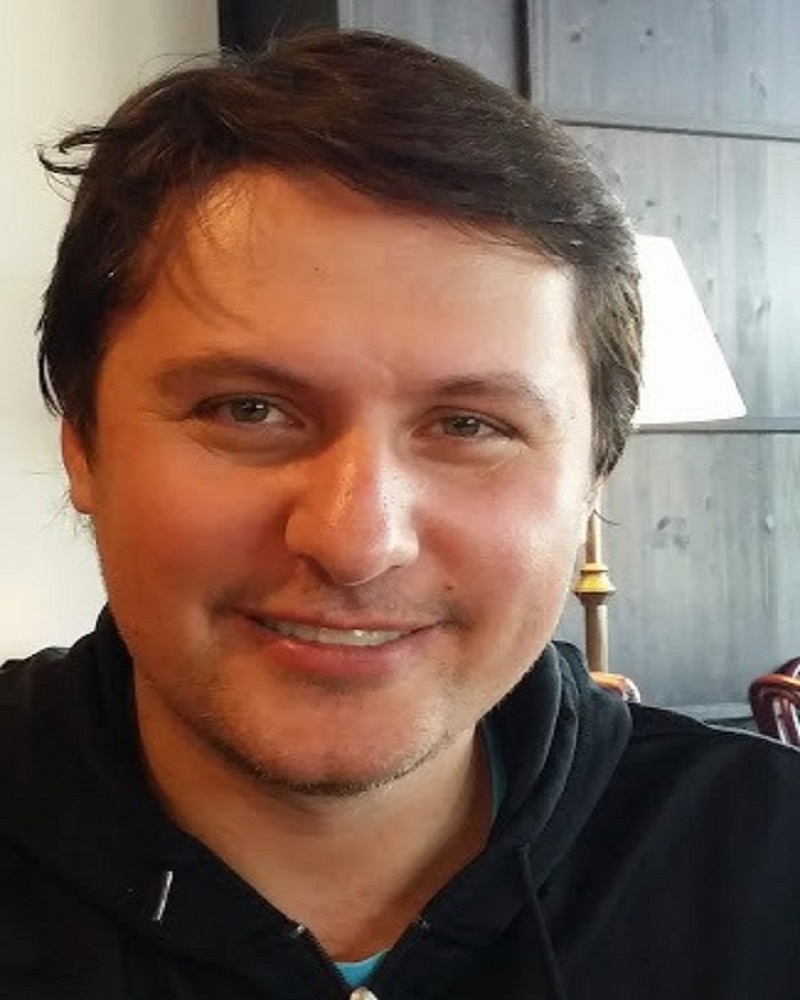}}]{Michael Sirivianos}
	is an Assistant Professor in Computer Engineering and Informatics at the Cyprus University of Technology.
	His current research interests lie in the fields of trust-aware design of distributed systems, device-centric authentication, federated identity management, discrimination based on personal data, cybersafety, suppression of false information in the social web, and transactional workload scalability.
\end{IEEEbiography}

\vskip -2\baselineskip plus -1fil
\begin{IEEEbiography}[{\includegraphics[width=1in,height=1.25in,clip,keepaspectratio]{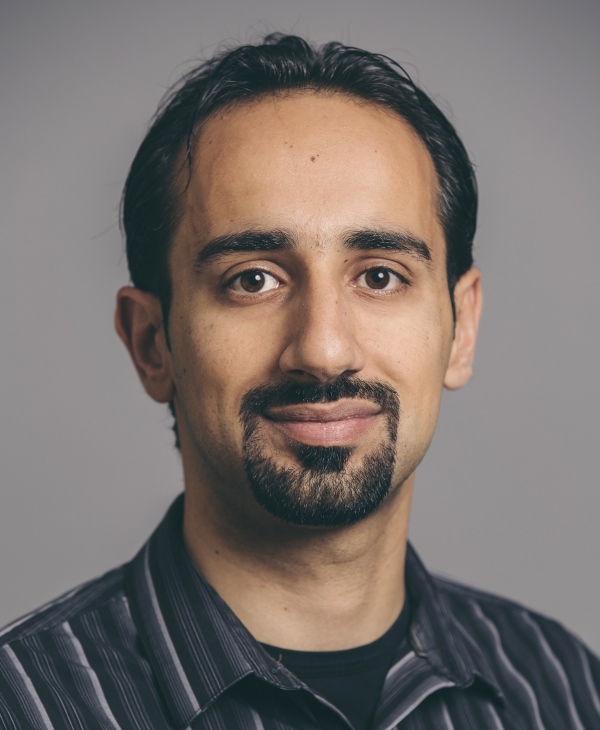}}]{Herodotos Herodotou}
	is an Assistant Professor in Computer Engineering and Informatics at the Cyprus University of Technology.
	His research interests are in large-scale data processing systems, database systems, and cloud computing. In particular, his work focuses on ease-of-use, manageability, and automated tuning of both centralized and distributed data-intensive computing systems.
\end{IEEEbiography}

\end{document}